\documentclass[onecolumn,showpacs,aps,prd]{revtex4}

\usepackage{latexsym}
\usepackage{amsmath,amssymb}

\newcommand{\lc}{\varepsilon}

\newcommand{\ds}{\displaystyle}
\newcommand{\del}{\partial}
\newcommand{\itGamma}{\mit{\Gamma}}

\newcommand{\tr}{\mathop{\rm tr}\nolimits}
\newcommand{\sgn}{\mathop{\rm sgn}\nolimits}
\renewcommand{\Re}{\mathop{\rm Re}\nolimits}
\renewcommand{\Im}{\mathop{\rm Im}\nolimits}

\newcommand{\prirodni}{\ensuremath{\mathbb{N}}}

\newcommand{\realni}{\ensuremath{\mathbb{R}}}
\newcommand{\kompleksni}{\ensuremath{\mathbb{C}}}
\newcommand{\poljeP}{\ensuremath{\mathbb{P}}}

\newcommand{\cD}{{\cal D}}

\newcommand{\cH}{{\cal H}}

\newcommand{\cM}{{\cal M}}
\newcommand{\cN}{{\cal N}}
\newcommand{\cO}{{\cal O}}

\begin{document}

\title{Cosine problem in EPRL/FK spinfoam model}

\author{Marko Vojinovi\'c}
\affiliation{Grupo de F\'isica Matem\'atica da Universidade de Lisboa \\ Av. Prof. Gama Pinto 2, 1649-003 Lisboa, Portugal \\ e-mail: {\tt vmarko@ipb.ac.rs}}


\begin{abstract}
We calculate the classical limit effective action of the EPRL/FK spinfoam model of quantum gravity coupled to matter fields. By employing the standard QFT background field method adapted to the spinfoam setting, we find that the model has many different classical effective actions. Most notably, these include the ordinary Einstein-Hilbert action coupled to matter, but also an action which describes antigravity. All those multiple classical limits appear as a consequence of the fact that the EPRL/FK vertex amplitude has cosine-like large spin asymptotics. We discuss some possible ways to eliminate the unwanted classical limits.
\end{abstract}

\maketitle

\section{\label{SectionIntroduction}Introduction}

One of the approaches to the problem of quantization of the gravitational field is called Loop Quantum Gravity. Historically it started as a canonical quantization of general relativity expressed in the language of the Ashtekar variables \cite{AstekarPrviRad}, and has since developed into a mainstream research direction \cite{RovelliBook}. Loosely speaking, it is split into the canonical quantization approach, today called {\em canonical} LQG, and the path integral approach, called {\em covariant} LQG. There are many concrete covariant formulations of LQG, and they are collectively called {\em spinfoam models}.

The main aim of every spinfoam model is to give a rigorous definition of the path integral for the gravitational field,
\begin{equation} \label{GravitacioniPathIntegral}
Z = \int \cD g_{\mu\nu} \exp \left( iS_{EH}[g_{\mu\nu}]\right) ,
\end{equation}
where $g_{\mu\nu}$ is the spacetime metric and $S_{EH}[g_{\mu\nu}]$ is the Einstein-Hilbert action for general relativity. This is done by discretization of the spacetime manifold in some particular way. One starts by considering a $2$-complex $\sigma$, consisting of vertices $v$, edges $e$ connecting the vertices, and faces $f$ bounded by vertices and edges. Each of these objects are suitably ``colored'' by certain variables $c$ which represent the gravitational degrees of freedom, and each are assigned an ``amplitude'' (some function of the colors), which establishes its contribution to the path integral. The colored $2$-complex $\sigma$ is called a spin foam, and the corresponding path integral is then defined in the following way:
$$
Z_{\sigma} = \sum_c \prod_{f\in\sigma} A_f(c) \prod_{e\in\sigma} A_e(c) \prod_{v\in\sigma} A_v(c).
$$
Here $A_f(c)$, $A_e(c)$ and $A_v(c)$ are the face amplitude, edge amplitude and the vertex amplitude, respectively. The colors are summed over if they take values in a discrete set, or integrated over if they take values in a continuum set. The $2$-complex may be dual to some triangulation $T(\cM)$ of some $4$-dimensional manifold $\cM$, and may have a boundary.

The colors are most often chosen such that on the boundary they correspond to the gravitational degrees of freedom of the canonical LQG formalism. The amplitudes should be chosen such that they have correct gluing properties along the boundary, that the whole state sum $Z$ be finite, and that in the classical limit the theory reduces to the ordinary general relativity.

There have been many spin foam models proposed in the literature, with varying degrees of success in satisfying the above criteria. The most successful model so far is called the EPRL/FK model \cite{EPRL,FK}, and it has been extensively studied in the literature. Its main ingredient is the particular choice of the vertex amplitude $A_v$, denoted $W_v$, which in the classical limit has the asymptotic form \cite{BarretEPRLasymptotics}
\begin{equation} \label{SkicaAsimptotikeVerteksa}
W_v \sim e^{iS_v} + e^{-iS_v} \sim \cos S_v,
\end{equation}
where $S_v$ is the area-Regge action for one $5$-simplex dual to the vertex $v$. If the vertex amplitude had asymptotics of the form $e^{iS_v}$, one could write the classical limit of the state sum in the form
$$
Z \sim \sum \prod_v e^{iS_v} \sim \sum e^{i\sum_v S_v} \sim \int e^{iS_R},
$$
where $S_R$ could be considered as the Regge action for the triangulation $T(\cM)$ of the manifold $\cM$ dual to the $2$-complex $\sigma$. In a suitable limit this could give a rigorous definition for the gravitational path integral (\ref{GravitacioniPathIntegral}), thus making the connection with the classical limit of the theory. Unfortunately, the vertex amplitude does not have asymptotics of the form $e^{iS_v}$, but rather of the form $\cos S_v$, which makes the above connection to (\ref{GravitacioniPathIntegral}) much more complicated. This is usually called the {\em cosine problem} in the literature.

In this paper we will try to provide a detailed analysis of the consequences of the asymptotics (\ref{SkicaAsimptotikeVerteksa}) for the classical limit of the EPRL/FK spin foam model. We will employ the background field method \cite{Abbott,KleinertHagen,MVeffAct,MVjpcs} for calculating the effective action of the theory in the classical limit, and analyze in detail the consequences of the conjugate term in (\ref{SkicaAsimptotikeVerteksa}). Our main result is that the EPRL/FK model, in its original form, indeed does have a correct classical limit, but also features additional classical limits, which do not correspond to general relativity. In particular, one of the classical limits predicts {\em antigravity}. We will also discuss some of the possible ways the EPRL/FK model could be modified in order to eliminate these extra classical limits.

It is important to note that various aspects of the effective action, in particular antigravity, become obvious and appreciable only after coupling the spin foam model to matter fields. Namely, the attractive or repulsive character of the gravitational field is dictated by the relative sign between the gravitational action and matter action. Thus, if matter is absent from the effective action, one cannot distinguish between gravity and antigravity. This is the main reason why the appearance of antigravity might seem surprising. Namely, the coupling of matter fields to spin foam models has been already considered in the literature (see for example \cite{RovelliMaterijaJedan,RovelliMaterijaDva,RovelliWilsonEwing,RovelliChristodoulouRiello} and references therein), but so far never in the context of the classical limit. As we shall see, antigravity is directly related to the minus sign in the second exponent in (\ref{SkicaAsimptotikeVerteksa}). In order to keep the discussion as general as possible, we shall not specify any particular properties of the matter fields, except for some generic properties of their classical limit. In particular, we will not specify the form of the matter action, and we will not fix any particular number or type of the matter fields present in the theory.

The layout of the paper is as follows. In section \ref{SectionEPRLFKmodel} we will give a short introduction to the EPRL/FK spin foam model, and discuss the results for the asymptotics of the vertex amplitude. In section \ref{SectionEffectiveAction} we will introduce the background field method for evaluating the effective action in quantum field theory. Some details of the method will be discussed, in particular the possibility of having multiple classical limits in the same theory. Then we will adapt the method to the spin foam setting, which gives us the possibility to evaluate the effective action for the EPRL/FK model in the classical limit. Section \ref{SectionClassicalLimit} is devoted to the computation of the effective action in the classical limit for the EPRL/FK spin foam model. It is split into five subsections, each of which deals with one of the aspects of the calculation --- from the definition of the classical limit in the spin foam setting, through all the details of the computation of the effective action, to taking the continuum limit of the resulting expression in order to obtain the classical general relativity coupled to matter fields. The main interpretation of the obtained results is then given in section \ref{SectAntigravity}, where the various classical limits are discussed in detail. The discussion focuses on the appearance of antigravity, and its connection to the conjugate exponent in the asymptotics of the vertex amplitude. We also devote attention to some of the possible scenarios to resolve the cosine problem, discussing their benefits and drawbacks. In section \ref{SectionConclusions} we give our concluding remarks. The Appendix is devoted to fixing the notation, and some technical results are spelled out in detail.

Throughout the paper we work in the natural system of units, $c=\hbar=1$ and $G=l_p^2$, with the $(-,+,+,+)$ signature for the Minkowski metric. For the remaining conventions and notation, see Appendix \ref{AppendixA}.

\section{\label{SectionEPRLFKmodel}EPRL/FK spin foam model}

We will give a brief overview of the EPRL/FK spin foam model \cite{EPRL,FK}, in particular its Lorentzian formulation. For the motivation and details see \cite{RovelliSpinFoamOverview} and references therein.

Given a fixed $2$-complex $\sigma$ with no boundary, denote its faces, edges and vertices as $f$, $e$, $v$, respectively. Assume that all vertices are $5$-valent. Each edge $e$ is connected to $4$ faces.

The assignment of color variables on this structure goes as follows. First, each face $f$ is colored by an irreducible representation $j_f\in \prirodni_0 /2$ of $SU(2)$. Second, in the Livine-Speziale coherent state basis \cite{LivineSpeziale}, one assigns a normalized vector $\vec{n}_{ef}\in S^2$ to each pair $ef$ describing one of the $4$ faces connected to a given edge $e$. In the EPRL/FK model there are no more labels, so that the gravitational degrees of freedom are described by the values of $j_f$ and $\vec{n}_{ef}$ for each face and edge on the $2$-complex $\sigma$. A $2$-complex colored in this way is called a spin foam.

Next, we introduce the choice of the face, edge and vertex amplitudes. The choice of face-amplitude is mainly determined by the gluing properties of spin foams with boundaries \cite{BianchiFaceAmplitude} and is given by
$$
A_f(j_f) = \dim j_f \equiv 2j_f + 1.
$$
Since the number of edges $E$ is proportional to the number of vertices $V$, $2E=5V$, one can always redefine the vertex amplitude $A_v(j_f,\vec{n}_{ef})$ such that all edge amplitudes are absorbed into the vertex amplitude. Consequently we can make a trivial choice for the edge amplitudes,
$$
A_e = 1.
$$
Finally, the vertex amplitude is the most complicated part of the EPRL/FK model. In the $(j,\vec{n})$ basis it can be expressed as
$$
\begin{array}{ccl}
A_v^{\rm EPRL/FK} & \equiv & \ds W_v(j,\vec{n}) = \cN \int_{SL(2,\kompleksni)^5} \prod_{a=1}^5 dg_a \, \delta(g_5) \\
 & & \ds
 \int_{(\kompleksni\poljeP^1)^{10}} \prod_{k=1}^{10} dz_k \, \Omega_k(g,z) \exp\left( \sum_{f'=1}^{10} j_{f'} \log w_{f'}(\vec{n},g,z) \right), \\
\end{array}
$$
where $w_f$ and $\Omega_k$ are some functions and $\cN$ is some normalization constant. See for example \cite{BarretEPRLasymptotics} for details.

Once the labels and amplitudes have been specified, we can write the state sum for the EPRL/FK spinfoam model as
\begin{equation} \label{DefinicijaStateSume}
Z_{\sigma} = \sum_j \int \prod_{ef} d\vec{n}_{ef} \prod_f \left(2 j_f + 1\right) \prod_v W_v(j,\vec{n}).
\end{equation}
Here the sum over $j$ means the sum over $j_f\in \prirodni_0 /2$ for all faces $f\in\sigma$, while the integral is over a sphere $S^2$ for every $ef$ pair belonging to $\sigma$. Similarly, products over $v$ and $ef$ range over all vertices and edge-face pairs in $\sigma$.

The EPRL/FK spinfoam model has numerous interesting properties. However, for our analysis two particular results will be important. The first is the physical interpretation of the $j_f$ variables. Namely, given a face $f$ and its label $j_f$, the area of a triangle dual to the face $f$ is
\begin{equation} \label{SpektarOperatoraPovrsine}
A_f = 8\pi\gamma l_p^2 \sqrt{j_f(j_f+1)}.
\end{equation}
Here $A_f$ is the area, $l_p$ is the Planck length, and $\gamma$ is the Barbero-Immirzi parameter. The second result is the asymptotic formula for the vertex amplitude $W_v$ in the limit where all spins $j$ are uniformly scaled to infinity, $j_f\to\infty$ (see \cite{BarretEPRLasymptotics}):
\begin{equation} \label{AsimptotkaVerteksAmplitude}
W_v(j,\vec{n}) \approx W_v^{\rm asymp}(j) \equiv N_+(j) e^{i\gamma S_v(j)} + N_-(j) e^{-i\gamma S_v(j)} + o(j^{-12}), \qquad (j\to\infty).
\end{equation}
See Appendix \ref{AppendixA} for the definitions of ``small-$o$'' and ``big-$O$'' symbols. Here $N_+(j)$ and $N_-(j)$ are some functions that scale as $O(j^{-12})$, while $S_v(j)$ is a function that scales as $O(j)$ and is given as
$$
S_v(j) = \sum_{f\in v} j_f \Theta_{v f}(j).
$$
Taking into account (\ref{SpektarOperatoraPovrsine}) in the limit $j\to\infty$, this function has the structure similar to the Regge action for a single $4$-simplex dual to the vertex $v$. Namely, the $O(1)$ function $\Theta_{vf}(j)$ has the geometric interpretation of the dihedral angle corresponding to the triangle dual to the face $f$ within the $4$-simplex dual to the vertex $v$. Moreover, the variable $j_f$ is proportional to the area of a triangle dual to the face $f$.

It is important to note that the asymptotic formula (\ref{AsimptotkaVerteksAmplitude}) is valid if the variables $(j,\vec{n})$ satisfy the so-called {\em Regge geometry} conditions. This means that the variables take certain values such that they describe a geometrical $4$-simplex embedded in a $4$-dimensional Minkowski geometry. There are also other possibilities --- for example the $(j,\vec{n})$ can be such that they describe a $4$-simplex embedded in a $4$-dimensional Euclidean geometry, or some other structure. In those cases the asymptotic formula is different, see \cite{BarretEPRLasymptotics} for details.

Finally, for what follows we need to generalize the EPRL/FK model in order to introduce matter fields. To this end, we denote all matter fields present in the theory as $\phi_r$, where $r$ counts all degrees of freedom of matter fields in the $2$-complex $\sigma$. The matter is coupled to gravity through the redefinition of the vertex amplitude
\begin{equation} \label{EPRLFKverteksAmplitudaSaMaterijom}
A_v (j_f,\vec{n}_{ef},\phi_r) = W_v(j_f,\vec{n}_{ef}) e^{iS_v^{\rm matter}(j_f,\vec{n}_{ef},\phi_r)},
\end{equation}
where $S_v^{\rm matter}(j,\vec{n},\phi)$ contains the details of the matter fields coupled to gravity, for a given $4$-simplex dual to the vertex $v$. The state sum for the theory with matter is
\begin{equation} \label{DefinicijaStateSumeSaMaterijom}
Z_{\sigma} = \sum_j \int \prod_{ef} d\vec{n}_{ef} \int \prod_r d\phi_r\; \prod_f \left[2 j_f + 1\right] \prod_v W_v(j,\vec{n}) e^{iS_v^{\rm matter}(j,\vec{n},\phi)}.
\end{equation}
The domain of integration over the fields $\phi$ ranges over the possible values that the fields can take.

The choice of the fields and their dynamics encoded in $S_v^{\rm matter}$ can vary wildly depending on the specific model. In particular, fermions are notoriously difficult to couple to gravity, even in the case of classical general relativity (for one possible way to couple Dirac fermions to the EPRL/FK model see \cite{RovelliMaterijaJedan,RovelliMaterijaDva}). However, for the purposes of this paper we will not need to specify any details about the fields or their dynamics, except some general properties which will be discussed later.

Also, it is important to note that the choice (\ref{EPRLFKverteksAmplitudaSaMaterijom}) for the vertex amplitude represents a natural way to couple matter fields to gravity, but it is by no means the only possible way. More complicated ways exist, and we will discuss one particular class of them in section \ref{SectAntigravity}. The reason why we consider (\ref{EPRLFKverteksAmplitudaSaMaterijom}) as the most natural choice is twofold. First, if one ``freezes-out'' the gravitational degrees of freedom in the state sum, in the continuum limit the state sum is supposed to take the form of the ordinary path integral of quantum field theory for the matter fields on that ``frozen'' background geometry. For example, one possible goal would be that this path integral reproduce the Standard Model of particle physics. The choice (\ref{EPRLFKverteksAmplitudaSaMaterijom}) for the matter coupling is the simplest one that can implement this requirement. Second, as we know from the theory of general relativity, the prescription for the coupling between gravity and matter is dictated by the equivalence principle. Therefore, we want the equivalence principle to hold at least in the classical limit of any theory of quantum gravity. As we shall demonstrate later, the coupling (\ref{EPRLFKverteksAmplitudaSaMaterijom}) is compatible with the equivalence principle. We will discuss this point in more detail in subsection \ref{SubSectMatterCoupling}, where we will compare and contrast the coupling (\ref{EPRLFKverteksAmplitudaSaMaterijom}) with another, more complicated choice of the coupling, and examine the compatibility of each of the two couplings with the equivalence principle.

\section{\label{SectionEffectiveAction}Effective action}

In quantum field theory, the background field method (see \cite{Abbott,KleinertHagen} for a review) provides one with an efficient way to calculate the effective action $\itGamma[\phi]$ as a solution of the following functional integrodifferential equation:
\begin{equation} \label{DefinicijaEffDejstvaUtp}
e^{i\itGamma[\phi]} = \int \cD \phi' e^{ i S[\phi+\phi'] - i\int \frac{\delta\itGamma[\phi]}{\delta\phi}\phi' }.
\end{equation}
Here $S[\phi]$ is the action of the classical theory, and $\phi$ is the arbitrary field configuration (the background) for which the effective action $\itGamma$ is to be calculated. In Appendix \ref{AppendixC} one can find an explicit derivation of this equation in the context of quantum field theory. It is very important to note that (\ref{DefinicijaEffDejstvaUtp}) represents an {\em off-shell} equation, in the sense that the background field $\phi$ is not assumed to satisfy any equations of motion, but is instead completely arbitrary (see Appendix \ref{AppendixC} for details).

It is usually not possible to find a general solution for the equation (\ref{DefinicijaEffDejstvaUtp}) explicitly. Instead, one must resort to various approximations. One of the most important approximations is the semiclassical limit.

The traditional way to establish a semiclassical limit is to take the limit where the quantum deformation parameter $\hbar $ goes to zero. Writing $\hbar$ explicitly in (\ref{DefinicijaEffDejstvaUtp}) amounts to substitutions
$$
\itGamma \to \frac{1}{\hbar} \itGamma, \qquad S \to \frac{1}{\hbar}S.
$$
However, in the natural system of units we have $c=\hbar =1$ and the reduced Planck constant is normalized to one. Given this normalization (which we use throughout the paper), the equivalent limit is to take $S[\phi] \to \infty$, which is usually achieved by taking $\phi \to\infty$. Namely, the semiclassical limit is by definition a physical configuration of the fields such that the action is $S[\phi]\gg \hbar \equiv 1$. In most cases the action is a polynomial functional of the fields, so in the semiclassical limit we can assume that both the background fields $\phi$ and the action $S[\phi]$ tend to infinity.

Therefore, in the case $S[\phi] \to\infty$, the equation (\ref{DefinicijaEffDejstvaUtp}) can be solved by an iterative asymptotic procedure. In particular, one writes the effective action as a sum of terms
\begin{equation} \label{AsimptotskiRazvojEffDejstva}
\itGamma[\phi] = \itGamma_0[\phi] + \itGamma_1[\phi] + \itGamma_2[\phi] + \dots ,
\end{equation}
where each consecutive term is assumed to be much smaller than the previous one in the asymptotic expansion,
$$
\itGamma_0 = O(S), \qquad \itGamma_{k+1} = o(\itGamma_k), \qquad \forall k\in\prirodni_0,
$$
where $S\to\infty$. Substituting the expansion of $\itGamma$ into (\ref{DefinicijaEffDejstvaUtp}) and calculating iteratively the $\itGamma_0$ and $\itGamma_1$ terms using the stationary point method, one obtains the following expression for the effective action,
$$
\itGamma[\phi] = S[\phi] + \frac{i}{2} \tr\log \frac{\delta^2S[\phi]}{\delta\phi\delta\phi} + o(\log S''),
$$
which has the asymptotic structure $\itGamma = O(S) + O(\log S) + o(\log S)$. The effective action is complex due to the Lorentzian nature of the path integral. After using the Wick rotation prescription (see \cite{MVeffAct} for details),
\begin{equation} \label{WickRotacija}
\itGamma_{\rm complex} \to \itGamma_{\rm real} = \Re(\itGamma_{\rm complex}) + \Im(\itGamma_{\rm complex}),
\end{equation}
one finally obtains the familiar expression for the effective action in the semiclassical limit,
$$
\itGamma[\phi] = S[\phi] + \frac{1}{2} \tr\log S''[\phi].
$$
For a review of the details of the calculation, see Appendix \ref{AppendixB}.

It is important to note that the equation (\ref{DefinicijaEffDejstvaUtp}) can have more than one solution in the classical limit. Depending on the structure of the action $S[\phi]$, it can become much larger than $\hbar$ in different ways for different configurations of the fields $\phi$. This will lead to multiple classical limits. For example, one can take two different configurations of fields, $\phi=\phi_1 \to \infty$ and $\phi=\phi_2 \to \infty$, such that the action $S[\phi]$ has different behavior in each limit, $S[\phi_1] = S_1[\phi_1] \to\infty$ and $S[\phi_2] = S_2[\phi_2]\to\infty$. In such cases one arrives at two different effective actions in the classical limit,
$$
\itGamma_1[\phi_1] = S_1[\phi_1] + o(S_1), \qquad \itGamma_2[\phi_2] = S_2[\phi_2] + o(S_2), \qquad \phi_1,\phi_2\to\infty.
$$
The two different effective actions describe the two different classical behaviors of the theory in two different physical regimes described by $\phi_1$ and $\phi_2$ respectively. In this sense, a theory can have more than one classical limit. This is of course not a problem, as long as $\phi_1$ and $\phi_2$ describe incompatible, mutually exclusive physical situations.

There can also be situations where (\ref{DefinicijaEffDejstvaUtp}) can give multiple solutions for the same configuration $\phi$ of the fields. For example, one can consider the action
$$
S[\phi] = -i \log \left[ e^{iS_1[\phi]} + e^{iS_2[\phi]} \right].
$$
Substituting this into (\ref{DefinicijaEffDejstvaUtp}) we obtain
$$
e^{i\itGamma[\phi]} = \int \cD \phi' \left[ e^{i S_1[\phi+\phi'] - i\int \frac{\delta\itGamma[\phi]}{\delta\phi}\phi' } + e^{ i S_2[\phi+\phi'] - i\int\frac{\delta\itGamma[\phi]}{\delta\phi}\phi'} \right].
$$
As we shall see in the next section, this equation has two solutions,
$$
\itGamma_1[\phi] = S_1[\phi] + o(S_1), \qquad \itGamma_2[\phi] = S_2[\phi] + o(S_2), 
$$
for the same physical configuration of the fields $\phi$, in the classical limit. There are situations where the two effective actions, although different, give equivalent classical equations of motion. For example, if $S_1[\phi] = a S_2[\phi] + b$, where $a$ and $b$ are constants. In this case, the two solutions are of course equivalent and represent the same classical limit.

However, it can also turn out that the two solutions produce inequivalent equations of motion. This last case is a problem for the theory --- if there is no unique classical limit for a given field configuration $\phi$, the theory does not have a well defined classical limit at all. Of course, this does not mean that the theory is inconsistent or otherwise ill-defined. From a mathematical point of view there is nothing wrong with the theory --- it only means that the classical limit, as defined by the limit $S[\phi] \gg \hbar$, does not exist. This is in complete analogy, for example, with a function $\sin (x)$ being well defined everywhere on a real line, while not having a well-defined value for $\ds\lim_{x\to\infty}\sin (x)$. On the other hand, from a physical point of view, this theory is problematic, since we expect that it has a unique classical limit. Namely, the existence and the uniqueness of the classical limit is a {\em physical requirement}, that ultimately comes from experiment.

In what follows, we shall be confronted with basically all these scenarios --- in the same theory we will find different solutions for $\itGamma$ for different configurations of $\phi$ (all being classical!), we will find different but equivalent solutions for the same choice of $\phi$, and finally different and inequivalent solutions for the same $\phi$. As we have elaborated, the first two scenarios do not pose a problem for the theory, while the third one does, albeit only from the standpoint of physics. 

Finally, we turn to an implementation of the equation (\ref{DefinicijaEffDejstvaUtp}) to the EPRL/FK spin foam model with matter fields. To this end, we define the effective action $\itGamma$ as a solution of the following integrodifferential equation (see also \cite{MVeffAct,MVjpcs}):
\begin{equation} \label{DefEffDejstvaUsf}
\begin{array}{ccl}
e^{i\itGamma(j,\vec{n},\phi)} & = & \ds \sum_{j'} \int \prod_{ef} d\vec{n}'_{ef} \int \prod_{r} d\phi'_r 
\; e^{-i\left( \sum_f\frac{\del \itGamma}{\del j_f}j'_f + \sum_{ef} \frac{\del\itGamma}{\del\vec{n}_{ef}} \vec{n}'_{ef} + \sum_{r} \frac{\del\itGamma}{\del \phi_r} \phi'_r  \right) } \\
 & & \ds \prod_f \left[2 \left(j_f+j'_f\right)+1\right] \prod_v W_v(j+j',\frac{\vec{n}+\vec{n}'}{\| \vec{n}+\vec{n}' \|}) e^{iS_v^{\rm matter}(j+j',\frac{\vec{n}+\vec{n}'}{\| \vec{n}+\vec{n}' \|},\phi+\phi')}. \\
\end{array}
\end{equation}
Equation (\ref{DefEffDejstvaUsf}) is a straightforward discretization of (\ref{DefinicijaEffDejstvaUtp}) for the spin foam setting, and it was first introduced in \cite{MVjpcs} for the pure gravity case. Note that the perturbation of the background fields $\vec{n}_{ef}$ had to be normalized, since both the background $\vec{n}$ and the perturbed background $\vec{n}+\vec{n}'$ must live on the unit sphere $S^2$. As in the case of the equation (\ref{DefinicijaEffDejstvaUtp}) itself, here we also emphasize that the background fields $j$, $\vec{n}$ and $\phi$ are {\em off-shell} --- they are not assumed to satisfy any equations of motion, but are instead completely arbitrary.

\section{\label{SectionClassicalLimit}Classical limit}

We will now employ the equation (\ref{DefEffDejstvaUsf}) to compute the effective action in the classical limit. Note that the result can depend on the particular configuration of the background fields, which means that there may be several different classical limits, corresponding to different choices of the background fields. As we have discussed in the previous section, this should not be surprising in any way.

We will study the classical limit in five steps. In the first step we will define what is meant by the classical limit in the spinfoam setting, and discuss the regime in which the fields behave classically. The second step will be the application of the asymptotic formula (\ref{AsimptotkaVerteksAmplitude}) and the discussion of the necessary assumptions. In the third step we will rearrange the equation (\ref{DefEffDejstvaUsf}) into a form suitable for the integration of $j'$ and $\phi'$ variables, and then perform the integration as a fourth step. Having obtained an effective action, the fifth step will be to analyze it and recast it into a continuum-variable language, thus obtaining the final form of the classical limit of the theory. It is convenient to split these five steps into five subsections.

\subsection{\label{SubSectClassicalLimit}Setting up the classical limit}

We begin by defining what field configurations we shall consider to be classical. First of all, we will restrict to the spin foam $2$-complex $\sigma$ which is dual to some triangulation $T(\cM)$ of some $4$-dimensional manifold $\cM$ with Lorentzian signature for the metric. Next, we will assume that the triangulation is very fine, so that the number of $4$-simplices $N_V$ is large, $N_V\gg 1$. Later we shall discuss further how large it should be. In addition, we will assume that the triangulation $T(\cM)$ is such that all triangles are spacelike, and consequently their areas are real numbers.

If the $2$-complex is dual to a triangulation, then it follows that there exist normals of the tetrahedra $\vec{n}_{ef}$ and spins $j_f$ which can be expressed as functions of edge lengths $L_\epsilon$, which define the metric in the triangulation $T(\cM)$. Therefore, we will consider the background $(j,\vec{n})$ which constitutes a Regge geometry. This means that $j=j(L)$ and $\vec{n} = \vec{n}(L)$. As a consequence, the tetrahedral closure constraint
\begin{equation} \label{ClosureConstraint}
\sum_{f\in v} j_f \vec{n}_{ef} = 0
\end{equation}
 is satisfied for every tetrahedron $e$ of a given $4$-simplex $v$. In addition, there are also other constraints present which enforce the Regge geometry across the whole manifold, as opposed to every $4$-simplex individually. Finally, the assumed Regge geometry of the triangulation implies that all vertices in the $2$-complex are $5$-valent, each corresponding to one $4$-simplex in $T(\cM)$.

At this point it is important to note that all these assumptions about geometry are not necessary for the classical limit, in the sense that the theory may have some classical limit even if these assumptions (or some of them) are not satisfied. As we have already noted, the theory may have several different classical limits simultaneously. That said, we are interested in this particular choice of the background geometry since that will eventually lead us to the classical Einstein-Hilbert action for the gravitational field, coupled to matter fields. Other choices of the background fields will lead to other types of classical limits, which will typically not be of Einstein-Hilbert type. In other words, one can relax any of the assumptions made above, and still (at least in principle) calculate the effective action $\itGamma$ from (\ref{DefEffDejstvaUsf}) in the limit when $j\to\infty$, $\phi\to\infty$ and the appropriate action $S[j,\vec{n},\phi]\to\infty$. However, in this paper we are not interested in those other classical limits, and rather restrict the background fields to the configurations discussed above. Just as an example, at the end of this section we will discuss shortly the effective action one can obtain by relaxing some of the assumptions of Regge geometry.

We now define the classical limit as the limit where edge lengths $L$ are much larger than the Planck length $l_p$. Given the relation (\ref{SpektarOperatoraPovrsine}) between the triangle area $A_f$ and the spin $j_f$ labeling the corresponding face $f$,
$$
A_f = 8\pi\gamma l_p^2 \sqrt{j_f(j_f+1)},
$$
the limit $L\gg l_p$ implies that $A \gg l_p^2$. Consequently we have
\begin{equation} \label{ClassicalLimitForArea}
\frac{1}{8\pi\gamma}\frac{A}{l_p^2} = \sqrt{j(j+1)} \approx j = O(N_j) \gg 1,
\end{equation}
for the fixed value of the Barbero-Immirzi parameter $\gamma$. Here we have introduced the number $N_j\gg 1$ which represents the scale for $j$. Later on we will discuss how large $N_j$ should be, along with the scale for the number of vertices $N_V$.

Regarding the matter field action $S_v^{\rm matter}$, we will need one additional assumption,
\begin{equation} \label{ZahtevZaMateriju}
\frac{\del}{\del \vec{n}_{ef}} \sum_v S_v^{\rm matter}(j,\vec{n},\phi) = o(j), \qquad j,\phi\to\infty,
\end{equation}
which tells us that the equations of motion of matter action with respect to the normal vectors $\vec{n}$ are automatically satisfied to the leading order. Physically, this means that matter fields are such that they do not ``feel'' the triangulation of spacetime, in the classical limit. Such an assumption is natural, since otherwise the measurements of matter fields in the classical limit would detect the piecewise linear structure of spacetime, which was not observed by any experiment so far. In the remainder of this paper, all calculations will be done up to $o(j)$ order, so given the assumption (\ref{ZahtevZaMateriju}), we can simply drop the dependence on $\vec{n}$ in the matter action and write
\begin{equation} \label{ZahtevZaMaterijuPrepisan}
S_v^{\rm matter}(j,\vec{n},\phi) = S_v^{\rm matter}(j,\phi) + o(j).
\end{equation}

All of the above preliminary considerations provide us with a setup to calculate the classical limit as the asymptotic limit for $\itGamma[j,\vec{n},\phi]$ when $j,\phi\to\infty$. Note that the normals $\vec{n}$ are normalized to live on a unit sphere $S^2$, and consequently they do not scale.

Now we turn to solving the equation (\ref{DefEffDejstvaUsf}) in the classical limit. We will be interested only in the leading order contribution to the effective action $\itGamma$, and we will ignore the subleading terms $\itGamma_1$, $\itGamma_2$, etc. in (\ref{AsimptotskiRazvojEffDejstva}).

\subsection{Approximating the vertex amplitude}

The first step is to discuss the application of the asymptotic formula (\ref{AsimptotkaVerteksAmplitude}) in the equation (\ref{DefEffDejstvaUsf}) for the effective action. Namely, according to (\ref{AsimptotkaVerteksAmplitude}), the vertex amplitude $W_v$ is peaked on the choice for $(j,\vec{n})$ which constitute the Regge geometry, and is suppressed otherwise. This means that in (\ref{DefEffDejstvaUsf}) each vertex amplitude
\begin{equation} \label{JedanPerturbovanVerteks}
W_v(j+j',\frac{\vec{n}+\vec{n}'}{\| \vec{n}+\vec{n}' \|})
\end{equation}
gives a leading order contribution in $j$ only for the choices of $\vec{n}'$ and $j'$ such that the combined variables $(j+j',\vec{n}+ \vec{n}')$ again constitute a Regge geometry. Otherwise, the vertex amplitude will be suppressed. Therefore, we can restrict the integration domain of $\vec{n}_{ef}'$ such that only Regge configurations appear in (\ref{JedanPerturbovanVerteks}). Then we can employ (\ref{AsimptotkaVerteksAmplitude}) and (\ref{ZahtevZaMaterijuPrepisan}) to rewrite (\ref{DefEffDejstvaUsf}) in the form
\begin{equation} \label{EffDejstvoIntN}
\begin{array}{ccl}
e^{i\itGamma(j,\vec{n},\phi)} & = & \ds \sum_{j'} \int \prod_{ef} d\vec{n}'_{ef} \int \prod_{r} d\phi'_r 
\; e^{-i\left( \sum_f\frac{\del \itGamma}{\del j_f}j'_f + \sum_{ef} \frac{\del\itGamma}{\del\vec{n}_{ef}} \vec{n}'_{ef} + \sum_{r} \frac{\del\itGamma}{\del \phi_r} \phi'_r  \right) } \\
 & & \ds \prod_f \left[2 \left(j_f+j'_f\right)+1\right] \prod_v W_v^{\rm asymp}(j+j') e^{iS_v^{\rm matter}(j+j',\phi+\phi')}. \\
\end{array}
\end{equation}
Note that for notational simplicity we will not explicitly denote the restricted domain of integration for $\vec{n}'$ in (\ref{EffDejstvoIntN}) and subsequent equations.

It is also important to note that the only place where $\vec{n}'$ variables appear in the integrand is in the exponent involving the derivative of $\itGamma$. In particular, $\vec{n}'$ variables do not appear in the asymptotic formula for the vertex amplitude due to the fact that (\ref{AsimptotkaVerteksAmplitude}) does not depend on $\vec{n}$ at all. In addition, the requirement (\ref{ZahtevZaMateriju}) for the matter action removes its dependence on $\vec{n}$ in the leading order. As a consequence of this, in the calculations that follow, the integration over $\vec{n}'$ variables will be largely irrelevant. In particular, by the end of subsection \ref{SubSectIntegracija} the dependence of the effective action $\itGamma$ on the background field $\vec{n}$ will turn out to be of subleading order. Nevertheless, it is important to note that all calculations are based on the assumption that the background variables $\vec{n}$ satisfy the conditions of Regge geometry. In that sense, the dependence of the effective action $\itGamma$ on $\vec{n}$ is similar to the dependence of the vertex asymptotic formula (\ref{AsimptotkaVerteksAmplitude}) --- the explicit formula for the effective action will not depend on $\vec{n}$ but it will be derived under the assumption that the background $\vec{n}$ (together with $j$) constitutes a Regge geometry.

\subsection{Rearranging the exponents}

Now we want to rewrite the integrand in (\ref{EffDejstvoIntN}) as a sum of exponents. To this end, note first that each vertex amplitude $W_v$ contains two exponential terms, conjugate to each other. The product over all vertices then gives
$$
\prod_v \sum_{\lc_v} N_{v\lc_v}(j+j') e^{i\lc_v\gamma S_v(j+j')} ,
$$
where $\lc_v\in\{ -1, +1\}$ for every $v$. Since we have $N_V$ vertices in the spin foam $2$-complex $\sigma$, the above expression can be rearranged as
$$
\sum_{\lc_1} \dots \sum_{\lc_{N_V}} \prod_v N_{v\lc_v}(j+j') e^{i\lc_v\gamma S_v(j+j')}.
$$
We are interested in two particular configurations of the $\lc_v$ terms. In particular, the choice $(++\dots +)$ and the choice $(--\dots -)$ deserve special attention, while all other intermediate choices will be called ``mixed terms'' and denoted as $\{ \lc \}^*$ in the sum. Therefore, we rewrite the above sum such that we separate it into the ``all positive'' piece, the ``all negative'' piece and the ``mixed'' piece:
\begin{equation} \label{UsputniOblikVerteksAmp}
\begin{array}{c}
\ds \left[ \prod_v N_{v+}(j+j') \right] \prod_v e^{i\gamma S_v(j+j')} +
\left[ \prod_v N_{v-}(j+j') \right] \prod_v e^{-i\gamma S_v(j+j')} + \\
\ds +\sum_{\{\lc\}^* } \prod_v N_{v\lc_v}(j+j') e^{i\lc_v\gamma S_v(j+j')}.\\
\end{array}
\end{equation}
Concentrate on the ``all positive'' term. The product of the exponents can be written as the exponent of the sum, which can then be rearranged in the following way:
$$
\begin{array}{ccl}
\ds i\gamma \sum_v S_v(j+j') & = & \ds
i\gamma \sum_v \sum_{f\in v} (j_f+j'_f)\Theta_{vf}(j+j')
=
i\gamma \sum_f (j_f+j'_f) \sum_{v\in f} \Theta_{vf}(j+j') = \\
 & = & \ds
i\gamma \sum_f (j_f+j'_f) \delta_f(j+j')
=
i\gamma S_R(j+j'). \\
\end{array}
$$
Here we have introduced the deficit angle $\delta_f$ for the face $f$ and the area-Regge action $S_R(j)$ for the manifold $\cM$ as
\begin{equation} \label{DefAngleDefAreaRegge}
\delta_f(j) = \sum_{v\in f} \Theta_{vf}(j), \qquad S_R(j) = \sum_f j_f \delta_f(j).
\end{equation}
Note that in Lorentz geometry the deficit angle $\delta_f$ is the sum of the dihedral angles $\Theta_{vf}$ over all vertices connected to the face. This is in contrast to the Euclidean geometry, where the deficit angle would be $2\pi$ minus the sum of dihedral angles.

The product of factors $N$ in the ``all positive'' term can be rewritten as
$$
\prod_v N_{v+}(j+j') = \prod_v \exp\log N_{v+}(j+j') = e^{\sum_v \log N_{v+}(j+j')}.
$$
One can perform analogous transformations to the ``all negative'' and ``mixed'' terms, and rewrite (\ref{UsputniOblikVerteksAmp}) in the form
\begin{equation} \label{UsputniOblikVerteksAmpDva}
\begin{array}{c}
\ds e^{i\gamma S_R(j+j') + \sum_v \log N_{v+}(j+j')} +
e^{-i\gamma S_R(j+j') + \sum_v \log N_{v-}(j+j')} + \vphantom{\sum_{\{\lc\}^* }} \\
\ds \sum_{\{\lc\}^* } e^{i \gamma \sum_v \lc_v S_v(j+j') + \sum_v \log N_{v\lc_v}(j+j')}, \\
\end{array}
\end{equation}
which can be substituted into (\ref{EffDejstvoIntN}). In addition, equation (\ref{EffDejstvoIntN}) features the product of the exponents of $S_v^{\rm matter}$ and the product of face amplitudes, both of which can be rewritten as single exponents,
\begin{equation} \label{UkupnoDejstvoMat}
\prod_v \exp \left[iS_v^{\rm matter}(j+j',\phi+\phi')\right] = e^{iS_M(j+j',\phi+\phi')},
\end{equation}
and
\begin{equation} \label{UkupnaFaceAmp}
\prod_f \left[2 \left(j_f+j'_f\right)+1\right] = e^{ \sum_f \log \left[2 \left(j_f+j'_f\right)+1\right]}.
\end{equation}
Here we have introduced the total matter action for the $2$-complex $\sigma$ as
$$
S_M(j,\phi) = \sum_v S_v^{\rm matter}(j,\phi).
$$
Finally, substituting (\ref{UsputniOblikVerteksAmpDva}), (\ref{UkupnoDejstvoMat}) and (\ref{UkupnaFaceAmp}) into (\ref{EffDejstvoIntN}), we obtain:
\begin{equation} \label{EffDejstvoSaExp}
\begin{array}{ccl}
e^{i\itGamma(j,\vec{n},\phi)} & = & \ds \int \prod_f dj'_f \int \prod_{ef} d\vec{n}'_{ef} \int \prod_{r} d\phi'_r \\
 & & \ds
\left[
e^{iA_+(j,\vec{n},\phi,j',\vec{n}',\phi')} + e^{iA_-(j,\vec{n},\phi,j',\vec{n}',\phi')} + \sum_{\{\lc\}^* } e^{iA_{\lc}(j,\vec{n},\phi,j',\vec{n}',\phi')}
\right], \\
\end{array}
\end{equation}
where we have approximated the sum over $j'$ with an integral via the Euler-Maclaurin formula. The exponents $A_+$, $A_-$ and $A_{\lc}$ are obtained by collecting together all exponent factors in the ``all positive'', ``all negative'' and ``mixed'' sums:
$$
\begin{array}{ccl}
A_+(j,\vec{n},\phi,j',\vec{n}',\phi') & = & \ds \gamma S_R(j+j') + S_M(j+j',\phi+\phi') + \vphantom{\sum_f\frac{\del \itGamma}{\del j_f}} \\
 & & \ds + \sum_f \log \left[2 \left(j_f+j'_f\right)+1\right] + \sum_v \log N_{v+}\left(j+j' \right) - \\ 
 & & \ds - \sum_f\frac{\del \itGamma}{\del j_f}j'_f - \sum_{ef} \frac{\del\itGamma}{\del\vec{n}_{ef}} \vec{n}_{ef}' - \sum_{r} \frac{\del\itGamma}{\del \phi_r} \phi'_r, \\
\end{array}
$$
$$
\begin{array}{ccl}
A_-(j,\vec{n},\phi,j',\vec{n}',\phi') & = & \ds - \gamma S_R(j+j') + S_M(j+j',\phi+\phi') + \vphantom{\sum_f\frac{\del \itGamma}{\del j_f}} \\
 & & \ds + \sum_f \log \left[2 \left(j_f+j'_f\right)+1\right] + \sum_v \log N_{v-}\left(j+j' \right) - \\ 
 & & \ds - \sum_f\frac{\del \itGamma}{\del j_f}j'_f - \sum_{ef} \frac{\del\itGamma}{\del\vec{n}_{ef}} \vec{n}_{ef}' - \sum_{r} \frac{\del\itGamma}{\del \phi_r} \phi'_r, \\
\end{array}
$$
and
$$
\begin{array}{ccl}
A_{\lc}(j,\vec{n},\phi,j',\vec{n}',\phi') & = & \ds \gamma \sum_v \lc_v S_v(j+j') + S_M(j+j',\phi+\phi') + \vphantom{\sum_f\frac{\del \itGamma}{\del j_f}} \\
 & & \ds + \sum_f \log \left[2 \left(j_f+j'_f\right)+1\right] + \sum_v \log N_{v\lc_v}\left(j+j' \right) - \\ 
 & & \ds - \sum_f\frac{\del \itGamma}{\del j_f}j'_f - \sum_{ef} \frac{\del\itGamma}{\del\vec{n}_{ef}} \vec{n}_{ef}' - \sum_{r} \frac{\del\itGamma}{\del \phi_r} \phi'_r. \\
\end{array}
$$
Note that there is one $A_{\lc}$ exponent for each choice $(\lc_{1},\dots,\lc_{N_V}) \in \{ \lc \}^*$, and that each two $A_{\lc}$ terms are mutually different in general. The notation $A_{\lc}$ does not explicitly distinguish between these, but is rather a catch-all shorthand which denotes any exponent of the ``mixed'' type. While this is a slight abuse of notation, it will not lead to any confusion.

The effective action equation in the form (\ref{EffDejstvoSaExp}) is suitable to integration by the methods of Appendix \ref{AppendixB}.

\subsection{\label{SubSectIntegracija}Integration around the large background}

We will solve equation (\ref{EffDejstvoSaExp}) by first evaluating the integral over $j'$, and after that the integral over $\phi'$. To this end, in the limit $j\to\infty$, we need to know the scaling of various terms in exponents $A_+$, $A_-$ and $A_{\lc}$. Some of the terms have obvious scaling:
$$
S_R(j+j') = O(j), \qquad S_{v}(j+j') = O(j), \qquad \itGamma(j) = O(j),
$$
$$
\log \left[2 \left(j_f+j'_f\right)+1\right] = O(\log j), \qquad  \log N_{v\lc_v}\left(j+j' \right) = O(\log j), \qquad \frac{\del \itGamma}{\del j_f} = O(1),
$$
while the terms involving $\del\itGamma / \del \vec{n}$ and $\del\itGamma / \del\phi$ do not depend on $j'$ and can be taken in front of the $j'$ integral. The nontrivial part is to analyze how the matter action $S_M$ scales with $j$, and how this scaling correlates with the scaling of the fields $\phi$. This can be determined by simple dimensional analysis, but needs to be done on a case-by-case basis, for each type of matter field separately. Note that we do not need to know the exact way the matter action is coupled to the spin foam variables $j$ and $\vec{n}$. Rather, we only need to know that large $j$ means that areas are much larger than $l_p^2$, see (\ref{ClassicalLimitForArea}). Then, for ordinary scalar, Dirac and gauge vector fields in flat Minkowski spacetime with metric $\eta = (-,+,+,+)$ we have
$$
\begin{array}{lcl}
S_{\rm scalar}[\varphi] & = & \ds \int d^4x \left[ \frac{1}{2} \left( \del_{\mu}\varphi \right) \left( \del^{\mu}\varphi \right) + \frac{1}{2} m^2 \varphi^2 \right], \\
S_{\rm Dirac}[\psi] & = & \ds \int d^4x \, \bar{\psi} \left( i\gamma^{\mu} \del_{\mu} - m \right) \psi, \\
S_{\rm vector}[A] & = & \ds \int d^4x \left[ \frac{1}{4}F_{\mu\nu}F^{\mu\nu} + \frac{1}{2}m^2A_{\mu}A^{\mu} \right], \qquad F_{\mu\nu} \equiv \del_{\mu}A_{\nu} - \del_{\nu}A_{\mu}. \\
\end{array}
$$
Given that $x$ scales as $O(j^{1/2})$, simple dimensional analysis yields that each of these actions will scale as $O(j)$ if
\begin{equation} \label{SkaliranjePoljaMaterije}
\varphi = O(1), \qquad \psi = O(j^{-1/4}), \qquad A_{\mu} = O(1), \qquad m = O(j^{-1/2}).
\end{equation}
Therefore, by scaling the matter fields in this way, we have
$$
S_M = O(j).
$$
Note that the matter fields will eventually also be scaled to infinity, since the classical limit is achieved by taking $j,\phi\to\infty$. Nevertheless, we first perform the $j'$ integration by taking $j\to\infty$ and keeping the scale of $\phi$ as is given in (\ref{SkaliranjePoljaMaterije}). After this is done, we will perform the remaining $\phi'$ integration in the limit $\phi\to\infty$.

From this point on, the integration of (\ref{EffDejstvoSaExp}) proceeds according to Appendix \ref{AppendixB}. We expand each of the exponents $A_+$, $A_-$ and $A_{\lc}$ into power series in $j'$ around the background $j$, discard terms higher than $(j')^3$, and choose $\del\itGamma / \del j_f$ to cancel the term linear in $j'$. Ignoring all subleading terms, the cancellation of leading order linear terms will happen for the exponents $A_+$, $A_-$ and $A_{\lc}$, if we choose respectively
\begin{equation} \label{RazniIzboriZaGamaNula}
\begin{array}{lcl}
\itGamma_+ (j,\vec{n},\phi) & = &\ds \gamma S_R(j) + S_M(j,\phi) + \tilde{\itGamma}_+(\vec{n},\phi) , \vphantom{\ds\int} \\
\itGamma_- (j,\vec{n},\phi) & = &\ds - \gamma S_R(j) + S_M(j,\phi) + \tilde{\itGamma}_-(\vec{n},\phi) , \vphantom{\ds\int} \\
\itGamma_{\lc} (j,\vec{n},\phi) & = &\ds \gamma \sum_v \lc_v S_v(j) + S_M(j,\phi) + \tilde{\itGamma}_{\lc}(\vec{n},\phi) . \vphantom{\ds\int} \\
\end{array}
\end{equation}
At this point we need to discuss the equation (\ref{EffDejstvoSaExp}). Namely, from the above choices for $\itGamma$ it is obvious that one can eliminate linear terms in only one of the exponents of (\ref{EffDejstvoSaExp}), which means that all others will be suppressed. This effect is well-known in the literature and is called the {\em suppression mechanism}. For example, see \cite{RovelliPropagator,BianchiSuppresionMechanism,BianchiPropagatorJedan,BianchiPropagatorDva} for its application in the calculation of the graviton propagator for the EPRL/FK spinfoam model.

From the behavior of the exponents, it is easy to see that equation (\ref{EffDejstvoSaExp}) has more than one solution, which means that the theory has more than one classical limit. Each choice for $\itGamma$ in (\ref{RazniIzboriZaGamaNula}) will lead to one particular classical effective action. None of the choices is preferred over the others, as they all correspond to the limit $j\to\infty$. In section \ref{SectAntigravity} we shall discuss the physical consequences of this situation. However, in order to complete the calculation of the effective action, we will choose one of the exponents, say $A_+$, and keep it as dominant, while others will be suppressed. This corresponds to the choice $\itGamma_+$ in (\ref{RazniIzboriZaGamaNula}). Effective actions coming from the other choices can be calculated in analogous manner, by repeating the steps of the $\itGamma_+$ calculation.

Substituting $\itGamma_+$ into (\ref{EffDejstvoSaExp}), only the first exponent will give the dominant contribution. Dropping all other exponents, the integral over $j'$ can be evaluated by following the steps of Appendix \ref{AppendixB}. This will give, in addition to the leading order $\itGamma_+=O(j)$, a term of the type $\tr\log S''$. This term is of order $O(\log j)$ which is subleading, and can be neglected in the leading order. After appropriate cancellations, equation (\ref{EffDejstvoSaExp}) reduces to:
\begin{equation} \label{EffDejstvoSaExpMat}
e^{iS_M(j,\phi) + i\tilde{\itGamma}_+(\vec{n},\phi)} = \int \prod_{ef} d\vec{n}'_{ef}  e^{-i\sum_{ef} \frac{\del \tilde{\itGamma}_+}{\del\vec{n}_{ef}}\vec{n}_{ef}'} \vphantom{\ds\int} \int \prod_r d\phi'_r e^{iS_M(j,\phi+\phi') - i\sum_r \frac{\del S_M}{\del\phi_r}\phi'_r - i\sum_r \frac{\del\tilde{\itGamma}_+}{\del\phi_r}\phi'_r}.
\end{equation}
Note that the second integral on the right-hand side is a discretization of the usual QFT path-integral (\ref{DefinicijaEffDejstvaUtp}) for the matter action. Now we proceed by solving the equation in the limit $\phi\to\infty$, which represents the classical limit in QFT. Integrating the matter fields, the action term $S_M$ is expanded in power series in $\phi'$ around the background $\phi$, the cubic and higher terms drop out, the constant term cancels the corresponding term on the left-hand side, and one is left with linear and quadratic terms in $\phi'$. The requirement that the linear terms vanish is
$$
\frac{\del \tilde{\itGamma}_+}{\del\phi_r} = 0,
$$
which means that $\tilde{\itGamma}_+(\vec{n},\phi) = \tilde{\itGamma}_+(\vec{n})$. The quadratic term gives rise to a $\tr\log S''$ subleading term, and can be dropped. Therefore, after the integration of the matter fields, equation (\ref{EffDejstvoSaExpMat}) reduces to
\begin{equation} \label{EffDejstvoSaExpN}
e^{i\tilde{\itGamma}_+(\vec{n})} = \int \prod_{ef} d\vec{n}'_{ef}  e^{-i\sum_{ef} \frac{\del \tilde{\itGamma}_+}{\del\vec{n}_{ef}}\vec{n}_{ef}'} \vphantom{\ds\int}.
\end{equation}
This is a functional integrodifferential equation for $\tilde{\itGamma}_+(\vec{n})$. Note that it does not depend on $j$ and $\phi$ variables but rather only on $\vec{n}$, which are of order $O(1)$. Thus, any solution of (\ref{EffDejstvoSaExpN}) will also be of that order, so we conclude:
$$
\tilde{\itGamma}_+(\vec{n}) = O(1).
$$
As this is subleading to the other terms in (\ref{RazniIzboriZaGamaNula}), it can be dropped. In the end we are left with the leading order effective action
\begin{equation} \label{EffDejstvoRezultatZaPlus}
\itGamma_+ (j,\phi) = \gamma S_R(j) + S_M(j,\phi) + o(j,\phi) , \qquad (j,\phi\to\infty).
\end{equation}

\subsection{Continuum limit}

The final step in the analysis of the effective action is the continuum limit. Namely, once we have the expression (\ref{EffDejstvoRezultatZaPlus}) for the effective action, we want to recast it in the familiar variables of classical field theory. In particular, we want to express the gravitational degrees of freedom via the tetrad fields $e^a{}_{\mu}(x)$, which live on a smooth manifold $\cM$. For this to happen, we should recall several assumptions made in subsection \ref{SubSectClassicalLimit}.

The first step is to remember that we have restricted to a configuration of the background fields $j,\vec{n}$ which satisfy Regge geometry on a triangulation $T(\cM)$ of the manifold $\cM$. This means that there is a choice of edge-lengths $L_{\epsilon}$ such that
\begin{equation} \label{ReggeGeometryBackgroundVariables}
j_f=j_f(L_{\epsilon}), \qquad \vec{n}_{ef}=\vec{n}_{ef}(L_{\varepsilon}).
\end{equation}
Substituting this into the effective action, it becomes the function of edge-lengths $L$:
$$
\itGamma_+ (L,\phi) = \gamma S_R(j(L)) + S_M(j(L),\phi) + o(L,\phi) , \qquad (L,\phi\to\infty).
$$
Next, if we remember (\ref{ClassicalLimitForArea}) and (\ref{DefAngleDefAreaRegge}), the area-Regge action $S_R(j(L))$ can be rewritten as
$$
\gamma S_R(j(L)) = \gamma \sum_f j_f(L) \delta_f(j(L)) = \frac{1}{8\pi l_p^2} \sum_f A_f(L) \delta_f(L) = \frac{1}{8\pi l_p^2} S_R(L),
$$
where we have introduced the length-Regge action in the usual way,
$$
S_R(L) = \sum_f A_f(L)\delta_f(L).
$$
Note that the Barbero-Immirzi parameter $\gamma$ has canceled out of the equation. Consequently we end up with the effective action living on a triangulation $T(\cM)$,
$$
\itGamma_+ (L,\phi) = \frac{1}{8\pi l_p^2} S_R(L) + S_M(j(L),\phi) + o(L,\phi) , \qquad (L,\phi\to\infty).
$$
The effective action features the usual Regge-gravity coupled to matter fields, with an appropriate coupling constant.

Finally, we can invoke the requirements that the triangulation be ``fine'', which means that the areas of triangles $l_p^2 N_j$ and the number of $4$-simplices $N_V$ are both assumed to be very large. More precisely, we need the limit analogous to the limit taken in classical theory of fluids. There, the fluid --- originally consisting of a large number of individual molecules --- is approximated as a continuum. The ``volume element of the fluid'' is thus defined to be big enough to contain a large number of molecules, while at the same time small enough to be considered infinitesimal compared to macroscopic fluid motion. In the same sense, we need to consider triangulation $T(\cM)$ such that all edge-lengths $L$ of the simplices are much larger than the Planck length $l_p$, but at the same time still much smaller than any observable distance, so that they can be considered infinitesimal. This restricts the scale $N_j$ as
$$
l_p \ll l_p \sqrt{N_j} \ll L_{\rm observable}.
$$
Next, the number of $4$-simplices $N_V$ must also be suitably large --- each simplex must be large enough to be made of edges of size $l_p \sqrt{N_j}$, while it must remain small enough so that the discrete structure of the manifold is experimentally invisible. In particular,
$$
N_V \sim \frac{{}^{(4)}V_{\rm observable}}{{}^{(4)}V_{\rm 4-simplex}} \sim \frac{{}^{(4)}V_{\rm observable}}{l_p^4 N_j^2} \gg 1.
$$
In addition, the equations of motion for the normals $\vec{n}$ must be identically satisfied. The Regge action does not depend on these variables, so it satisfies this requirement. However, the matter action $S_M$ can in principle give nontrivial equations of motion. The requirement that this does not happen is the statement of assumption (\ref{ZahtevZaMateriju}).

Under such circumstances, one can substitute the edge-lengths $L$ with the tetrad field $e^a{}_{\mu}(x)$ living on the manifold $\cM$. The Regge action then gets transformed into the Einstein-Cartan action
$$
S_R(L) \to \frac{1}{2} S_{EC}[e] \equiv \frac{1}{2} \int \tr \star (e\wedge e) \wedge R(e).
$$
The matter action is also suitably converted, $S_M(L,\phi) \to S_M[e,\phi]$, so in the end taking the continuum limit amounts to writing the effective action in the form
\begin{equation} \label{EffDejstvoPlusKrajnjiRezultat}
\itGamma_+ [e,\phi] = \frac{1}{16\pi l_p^2} S_{EC}[e] + S_M[e,\phi].
\end{equation}
This concludes the analysis of the classical limit for the choice $\itGamma_+$ in (\ref{RazniIzboriZaGamaNula}).

As a final comment, let us consider a configuration of the background fields $j,\vec{n}$ such that the Regge geometry condition (\ref{ReggeGeometryBackgroundVariables}) is relaxed, while we still assume that the background normals $\vec{n}$ satisfy the Regge geometry constraint (\ref{ClosureConstraint}) for each vertex individually, so that the asymptotics (\ref{AsimptotkaVerteksAmplitude}) still holds. In that case one can again obtain an effective action in the classical limit, in the form (\ref{EffDejstvoRezultatZaPlus}):
\begin{equation} \label{EffDejstvoAreaReggePlusMaterija}
\itGamma_+ (j,\phi) = \gamma \sum_f j_f \delta_f(j) + S_M(j,\phi).
\end{equation}
However, due to the lack of conditions (\ref{ReggeGeometryBackgroundVariables}), this action is very different than the one discussed so far. In particular, the first term is the area-Regge action, rather than the usual length-Regge action. Namely, since the assumed background geometry does not necessarily correspond to a manifold triangulation $T(\cM)$, one cannot introduce the concept of edges with well-defined lengths in this geometry. Consequently, one must keep using the areas $j$ as variables for this action, which has nontrivial consequences. The variation with respect to $j$ gives the following equation of motion:
$$
\delta_f(j) + \frac{\del S_M}{\del j_f} = 0.
$$
In the absence of matter, this reduces to $\delta_f(j) = 0$, suggesting that the only vacuum solution is the flat space. However, it is not obvious that the vanishing of the deficit angle implies that curvature is zero, since in this kind of area-defined geometry the relation between the deficit angle and the parallel transport around a closed loop might be more complicated than naively expected. But in any case, the equations of motion need not have any resemblance to the Einstein equations, and can give completely different predictions.

The action (\ref{EffDejstvoAreaReggePlusMaterija}) is an example of one possible classical limit which is present in the theory, but very different from the expected Einstein-Hilbert action. As we have noted before, this is a consequence of the fact that equation (\ref{DefEffDejstvaUsf}) can have many different solutions, even in the limit $\hbar\to 0$. Given that the chosen background fields do not describe a Regge geometry, the action (\ref{EffDejstvoAreaReggePlusMaterija}) corresponds to a physical regime which is very different from the regime in which one can expect to obtain the Einstein-Hilbert action.

The fact that the effective action (\ref{EffDejstvoAreaReggePlusMaterija}) is not equivalent to the Einstein-Hilbert action does not in any way represent a problem for the theory. As we have argued in section \ref{SectionEffectiveAction}, it is quite possible, and even intuitively expected, that the same theory can have different classical limits in different physical regimes. In contrast to this, it can also happen that one obtains multiple inequivalent effective actions even for exactly the same physical regime, i.e. for the same choice of the background fields $j,\vec{n}$. Such a situation would indeed represent a problem for the theory, since the classical limit should be unique in any given physical regime. We will be confronted with this problem in the next section.

\section{\label{SectAntigravity}Conjugate exponent and antigravity}

In the previous section we have argued that the fact that effective action equation (\ref{DefinicijaEffDejstvaUtp}) or (\ref{DefEffDejstvaUsf}) can have many solutions in the classical limit is not a problem in itself, as long as the different limits correspond to different configurations of the background fields. This means that the theory has different classical behaviors in physically different situations. In this sense the effective actions given by (\ref{EffDejstvoPlusKrajnjiRezultat}) and (\ref{EffDejstvoAreaReggePlusMaterija}) correspond to different physical situations (presence and absence of Regge-geometry field configurations). However, if a theory has more than one classical limit for the {\em exact same field configuration}, then one has a problem, since there is no way to distinguish which effective action gives a correct description of the physics in the given regime.

Therefore, after obtaining the result (\ref{EffDejstvoPlusKrajnjiRezultat}), the most immediate question is what happens if we choose to compute the effective action around one of the other exponents in (\ref{EffDejstvoSaExp}). Given that each exponent is singled out by a suitable choice of $\itGamma$ in (\ref{RazniIzboriZaGamaNula}), we can choose $\itGamma_-$ instead of $\itGamma_+$. The only difference between these two actions is the minus sign in front of the area-Regge action. One can then recalculate the effective action all the way to the end, and instead of (\ref{EffDejstvoPlusKrajnjiRezultat}) obtain the following effective action:
\begin{equation} \label{EffDejstvoMinusKrajnjiRezultat}
\itGamma_- [e,\phi] = -\frac{1}{16\pi l_p^2} S_{EC}[e] + S_M[e,\phi].
\end{equation}
It is important to note that all the choices that we have made and properties of the background fields that we have used in the derivation remain exactly the same for both $\itGamma_+[e,\phi]$ and $\itGamma_-[e,\phi]$. Therefore, the theory provides us with two different classical limits which should be valid in exactly the same physical regime, for the same choice of the background fields.

The difference between the two actions is rather obvious --- the $\itGamma_-$ action has the ``wrong'' sign for the gravitational sector of the theory. In order to see what this means, one can vary the action $\itGamma_+$ with respect to the tetrad variables, and obtain the Einstein equations
$$
R_{\mu\nu} - \frac{1}{2}g_{\mu\nu}R = 8\pi l_p^2\; T_{\mu\nu},
$$
where $T_{\mu\nu}$ is the stress-energy tensor for the matter fields. In contrast, performing the same procedure for the action $\itGamma_-$ gives
$$
R_{\mu\nu} - \frac{1}{2}g_{\mu\nu}R = - 8\pi l_p^2\; T_{\mu\nu},
$$
where both the curvature and stress-energy tensors have the same form as in the previous case. One can be even more blunt and take the nonrelativistic approximation of these equations, and in a suitable limit derive the two versions of Newton's law of universal gravity:
$$
m\frac{d^2\vec{r}}{dt^2} = -l_p^2 \frac{mM}{r^2}\vec{e}_r, \qquad m\frac{d^2\vec{r}}{dt^2} = +l_p^2 \frac{mM}{r^2}\vec{e}_r.
$$
The only difference is in the sign of the gravitational constant $l_p^2$, which traces back all the way to the sign in front of the area-Regge term in (\ref{RazniIzboriZaGamaNula}). The physical interpretation is obvious --- one action, say $\itGamma_+$, predicts attractive gravity, while the other, $\itGamma_-$, predicts repulsive gravity, i.e. antigravity. Which action predicts one or the other depends on the details of the matter action $S_M$, but the key insight is that {\em both} limits are present in the theory.

The prediction of antigravity can be a serious problem for the EPRL/FK model. As we have argued above, one does not have a problem with all other possible classical limits of the theory, where any of our assumptions about the choice of the background fields are not satisfied. The point of the derivation of (\ref{EffDejstvoPlusKrajnjiRezultat}) was only to demonstrate that the expected classical limit (i.e. classical general relativity with matter) is indeed present among all those possible classical limits of the theory. However, the procedure has shown that along with this expected classical limit one can also find the antigravitational classical limit, valid in the very same physical regime. In addition to those, one can also find a whole sequence of ``mixed'' classical limits, obtained from $\itGamma_{\lc}$ in (\ref{RazniIzboriZaGamaNula}), which ``interpolate'' between the gravitational and antigravitational limits. These mixed effective actions do not even have a good continuum limit, since in those cases the gravitational constant changes sign arbitrarily from simplex to simplex, which would lead to gravitational force being either attractive or repulsive, depending on the point of the manifold. These classical limits also represent a serious problem for the theory, on the same footing as the antigravitational limit.

In order to resolve these issues, one needs to somehow eliminate those extraneous solutions for the effective action from the theory, while retaining the solution (\ref{EffDejstvoPlusKrajnjiRezultat}). This is extremely tricky to do, in particular because of the fact that there is no argument which could favor $\itGamma_+$ over $\itGamma_-$, since the two actions are completely equal in all respects, bar the sign of the gravitational constant.

Therefore, it appears necessary to modify the EPRL/FK model in some way, so that the problem of multiple classical limits does not happen. To that end, it is instructive to look how did multiple classical limits appear in the first place. First of all, they can be traced back to the sum of exponents in (\ref{EffDejstvoSaExp}). Each exponent is responsible for one choice of the effective action in (\ref{RazniIzboriZaGamaNula}). Tracing back further, these multiple exponents appear as a generic consequence of the fact that the EPRL/FK vertex amplitude $W_v$ has a cosine-like asymptotic expansion (\ref{AsimptotkaVerteksAmplitude}). The two conjugate exponents for a single vertex are multiplied in all possible combinations across all vertices, giving rise to multiple exponential terms in (\ref{EffDejstvoSaExp}).

It is important to stress that the asymptotics of the EPRL/FK vertex amplitude features both exponents, irrespective of the choice of fundamental variables used to describe the gravitational degrees of freedom in the theory. Namely, in section \ref{SectionEPRLFKmodel} we have formulated the EPRL/FK state sum (\ref{DefinicijaStateSume}) in terms of the Livine-Speziale coherent basis variables $(j,\vec{n})$, see \cite{LivineSpeziale}. This is by no means a unique choice --- one could instead choose the area-volume variables, or spin-intertwiner-holonomy variables (the first order formalism), or any other convenient set of variables that can be found in the literature. All these different choices of variables correspond to different choices of basis vectors in the kinematical Hilbert space of the theory. In that sense, they are all equivalent --- switching from one set of basis vectors to another amounts to rewriting the state sum in terms of one or the other set of variables.

In light of the problem of multiple classical limits, the question one could then ask is whether the presence of the two exponents in (\ref{AsimptotkaVerteksAmplitude}) is an intrinsic property of the vertex, or the artifact of the particular choice of variables. The origin of the two exponents in the asymptotics (\ref{AsimptotkaVerteksAmplitude}) has been studied in detail in \cite{JohnEngleJedan,JohnEngleDva,JohnEngleTri}. There it was established that the appearance of the two exponents is indeed an intrinsic property of the vertex amplitude, and cannot be eliminated by a change of basis. Namely, one constructs the vertex amplitude $W_v$ by the spinfoam quantization of the Holst action, which features a determinant of the tetrad fields. In contrast to the metric formulation of gravity, this determinant can be either positive or negative, depending on the tetrad fields at a given point in spacetime. The quantization procedure does not prefer either sign over the other, and therefore in the asymptotics (\ref{AsimptotkaVerteksAmplitude}) of $W_v$ both of these two signs give contributions in the form of two exponents of the Regge action. Therefore, the origin of these two exponents is completely independent of the basis variables, and one cannot eliminate one of the exponents by a simple change of basis.

We should also emphasize that there might be some ambiguity regarding what is considered to be a given physical regime in which the theory is supposed to have a unique classical limit. Namely, we can choose to express the theory in some suitable set of variables (denote them collectively as $K$) which is extended with additional degrees of freedom such that the choice of background fields automatically specifies only one exponent in the vertex amplitude (say, $K=K_+$), while the other exponent would correspond to some different choice of the background fields (say, $K=K_-$). In such $K$ variables, one could consider the second exponent as a different sector of the theory, which is expected to have a different classical limit, much like the limit of the non-Regge geometry (\ref{EffDejstvoAreaReggePlusMaterija}) discussed at the end of the previous section. In other words, the question is whether the two choices $K_+,K_-$ of the background fields correspond to the ``same physical regime'' (where the theory is supposed to have the same classical limit), or to two different physical regimes (where the theory is allowed to have different classical limits). As long as one is dealing with the EPRL/FK model, both of these backgrounds $K_+$ and $K_-$ expressed in these generalized variables must map to the same set of $(j,\vec{n})$ variables by integrating out the additional degrees of freedom. In this sense they must both correspond to the same physical regime. It is possible, however, to consider the full set of $K$-variables as fundamental, in which case one should interpret backgrounds $K_+$ and $K_-$ as physically different. In that case, though, one is generalizing the EPRL/FK model beyond its original set of variables. The analysis of such generalized models, while certainly very interesting and important in its own right, is out of the scope of this paper.

The conclusion of the whole analysis is that one needs to somehow modify the original theory, in order to eliminate the unwanted effective actions. There are several possible scenarios by which one could proceed in doing this. Each scenario has its own degree of success and its own shortcomings. We will discuss three of them.

\subsection{Redefining the vertex amplitude}

One obvious way to deal with this situation is to change the definition of the vertex amplitude $W_v$, such that in the large-spin limits it gives only one exponent. This has been suggested in \cite{MVeffAct,MVwavefuncassymptotics,JohnEngleJedan,JohnEngleDva,JohnEngleTri}. One way is to introduce the new vertex amplitude, $A_v(j,\vec{n})$ as
$$
A_v(j,\vec{n}) = \frac{1}{2N_+(j)} \left( W_v(j,\vec{n}) + \sqrt{W^2_v(j,\vec{n}) - 4 N_+(j)N_-(j)} \right),
$$
and write the spin foam state sum (\ref{DefinicijaStateSume}) with $A_v$ in place of $W_v$. Another possible way would be to define $A_v$ as
$$
A_v(j,\vec{n}) = \frac{N_+(j) W_v(j,\vec{n}) - N_-(j) W^*_v(j,\vec{n})}{N^2_+(j) - N^2_-(j)},
$$
where $W^*_v$ stands for the complex-conjugate of $W_v$. A third, more geometrical way to define $A_v$ has been proposed in \cite{JohnEngleDva,JohnEngleTri}. It is straightforward to verify for each of these definitions that in the limit $j\to\infty$ the amplitude $A_v$ has asymptotic behavior
$$
A_v \approx e^{i\gamma S_v(j)}.
$$
Repeating the calculation of the effective action, one sees that all exponents in (\ref{EffDejstvoSaExp}) vanish, except the first one which gives rise to (\ref{EffDejstvoRezultatZaPlus}) and consequently (\ref{EffDejstvoPlusKrajnjiRezultat}) as the only solution in the given regime. Therefore, eliminating the conjugate exponent from the definition of the vertex amplitude will certainly also eliminate all extraneous solutions for the effective action, including the antigravitational one.

The downside of this approach is that by redefining the vertex one is changing the theory, i.e. the calculated effective action does not correspond to the EPRL/FK model anymore. In addition, there is no unique way to redefine the vertex, which gives rise to multiple different models of quantum gravity. Each model has a correct classical behavior, but different quantum corrections. In absence of any experimental constraints, there is no way to prefer one model over the others.

\subsection{The boundary wavefunction}

Another method to eliminate the extra exponent has been introduced in \cite{RovelliModesto,RovelliPropagator}, see \cite{BianchiSuppresionMechanism} for a review. The idea is to calculate the spin foam path integral on a compact region of spacetime, with the boundary conditions embodied in the boundary wavefunction $\psi(j,\vec{n})$. The wavefunction represents the evaluation of the path integral outside the compact region, depends only on $j$ and $\vec{n}$ variables which intersect the boundary of the compact region, and is assumed to have the asymptotic form
$$
\psi(j,\vec{n}) \approx \psi_0(j) e^{i\sum_f j_f\theta_f}
$$
in the limit $j\to\infty$. Here $\psi_0$ is some real function (typically a Gaussian) while the $\theta_f$ are constants. In that setting one calculates the expectation value of any given observable $\cO(j,\vec{n})$ (which is defined only inside the compact region) as
$$
\left< \cO \right> = \frac{\ds \sum_j \int d\vec{n} \; \psi(j,\vec{n}) \cO(j,\vec{n}) \prod_f \left[ 2j_f+1 \right] \prod_v W_v }{\ds\sum_j \int d\vec{n} \; \psi(j,\vec{n}) \prod_f \left[ 2j_f+1 \right] \prod_v W_v}.
$$
The denominator must be different from zero, so using a convenient normalization of $\psi$, one requires it to satisfy the condition
$$
\sum_j \int d\vec{n} \; \psi(j,\vec{n}) \prod_f \left[ 2j_f+1 \right] \prod_v W_v = 1.
$$
This condition is equivalent to the requirement that $\psi$ is an element of the physical Hilbert space (the space of solutions of the theory), i.e. that it satisfies the Hamiltonian constraint
\begin{equation} \label{HamiltonianConstraint}
\hat{\cH} \psi = 0.
\end{equation}
Given this setup, one can calculate the expectation values of various observables in the limit $j\to\infty$, and use the coefficients $\theta_f$ to preselect the exponent one wants, while suppressing others. This is done in the same way as we did in solving (\ref{EffDejstvoSaExp}) by choosing one of the possible effective actions listed in (\ref{RazniIzboriZaGamaNula}). 

One could in principle try to implement this strategy when solving (\ref{DefEffDejstvaUsf}) to exclude all unwanted exponents and solve for the effective action (\ref{EffDejstvoPlusKrajnjiRezultat}) uniquely. However, there are several shortcomings to this approach. First, in this way one can obtain the effective action only inside some compact region of spacetime, which immediately raises the questions of the physical interpretation of such effective action. Namely, it is not obvious how could the equation (\ref{DefinicijaEffDejstvaUtp}) be rewritten to accommodate the boundary wavefunction $\psi$. And even if this is somehow done, it is not obvious what would it mean to have boundary conditions on the effective action. Second, one needs to verify that the boundary wavefunction $\psi$ with a given choices for the constants $\theta_f$ indeed satisfies the Hamiltonian constraint, which might not be the case. And third, even if one does establish that $\psi$ satisfies (\ref{HamiltonianConstraint}), this would imply that the complex-conjugate function $\psi^*$ also satisfies the same constraint:
$$
\hat{\cH} \psi^* = \left( \hat{\cH}^{\dagger} \psi \right)^* = \left( \hat{\cH} \psi \right)^* = 0.
$$
This is due to the fact that the Hamiltonian is a self-adjoint operator, $\hat{\cH}^{\dagger} = \hat{\cH}$. Given that $\psi^*$ is also a valid boundary wavefunction, one can use it instead of $\psi$ to solve for the effective action. Following through the suppression mechanism, one sees that the only non-suppressed exponent will now be the ``wrong'' one, giving rise to the antigravitational effective action (\ref{EffDejstvoMinusKrajnjiRezultat}). Thus, one sees that both effective actions are still present in the theory.

The only way to circumvent this would be to claim that the Hamiltonian is not self-adjoint, in order to eliminate $\psi^*$ from the set of solutions, while retaining $\psi$. In other words, one has to prove that both equations
$$
\hat{\cH} \psi = 0, \qquad \hat{\cH} \psi^* \neq 0,
$$
are satisfied simultaneously. However, even if one somehow manages to prove these two equations, the cost is very high --- the EPRL/FK model would then explicitly violate unitarity.

Therefore, while the upside of this method is that one stays within the EPRL/FK model, there are numerous downsides --- the effective action can be evaluated only inside some compact region of spacetime, the Hamiltonian constraint must be satisfied for the boundary wavefunction $\psi$, and the theory must be non-unitary, in order to eliminate $\psi^*$ from the set of solutions.

\subsection{\label{SubSectMatterCoupling}Redefining the matter coupling}

The third way to eliminate the unwanted exponent is to redefine the way matter couples to the EPRL/FK vertex amplitude. Namely, in contrast to (\ref{EPRLFKverteksAmplitudaSaMaterijom}), one could in principle construct the vertex amplitude $A_v(j,\vec{n},\phi)$ such that it reduces to the EPRL/FK vertex amplitude $W_v$ when matter fields are in their vacuum state (say, $\phi = 0$),
$$
A_v(j,\vec{n},0) = W_v(j,\vec{n}),
$$
and that in the limit $j\to\infty$ (and possibly $\phi\to\infty$) it has the asymptotics
\begin{equation} \label{NovoKuplovanjeMaterije}
A_v(j,\vec{n},\phi) \approx N_+(j,\phi) e^{i\gamma S_v(j) + i S_v^{\rm matter}(j,\phi)} + N_-(j,\phi) e^{-i\gamma S_v(j) - i S_v^{\rm matter}(j,\phi)}.
\end{equation}
Note that now in the conjugate exponent the sign is flipped not only for the gravitational sector, but also for the matter sector, so that the relative sign always stays the same.

Repeating the calculation of the effective action, one would obtain the following solutions for the effective action (compare with (\ref{RazniIzboriZaGamaNula}))
\begin{equation} \label{RazniIzboriZaGamaNulaZaNoviVerteks}
\begin{array}{lcl}
\itGamma_+ (j,\phi) & = &\ds \gamma S_R(j) + S_M(j,\phi) , \vphantom{\ds\int} \\
\itGamma_- (j,\phi) & = &\ds - \gamma S_R(j) - S_M(j,\phi) , \vphantom{\ds\int} \\
\itGamma_{\lc} (j,\phi) & = &\ds \sum_v \lc_v \left[ \gamma S_v(j) + S_v^{\rm matter}(j,\phi) \right] . \vphantom{\ds\int} \\
\end{array}
\end{equation}
In the continuum limit, the $\itGamma_+$ solution would lead to the expected classical action (\ref{EffDejstvoPlusKrajnjiRezultat}), and the $\itGamma_-$ solution would lead to the same result (\ref{EffDejstvoPlusKrajnjiRezultat}) up to an overall minus sign. However, this overall minus sign is classically unobservable, and thus does not lead to antigravity.

The main problem with this approach is the fact that it fails to eliminate the $\itGamma_{\lc}$ solutions. Moreover, not only that they do not give the correct effective action, but they even violate the equivalence principle, in the following way.

Informally stated, the equivalence principle says that in local Minkowski coordinates, all laws of physics must reduce to their special-relativistic form. As a consequence, if spacetime is globally flat, i.e. the Minkowski manifold $M_4$, the effective action should take its special-relativistic form for the matter fields. Thus, to test if a given action complies with the equivalence principle, we may perform the following procedure. Given an action which describes gravity and some matter fields coupled to it in some way, one can ``freeze-out'' the gravitational degrees of freedom by evaluating the action on a flat Minkowski background geometry. The remaining action is the action for matter fields only. If this action is not equivalent to its special-relativistic form, the equivalence principle is violated.

It can be instructive to demonstrate this on a simple example. Consider the Ein\-st\-ein-Hilbert gravity with the nonzero cosmological constant $\Lambda$ and one real scalar matter field $\varphi$, described by the action
$$
S[g,\varphi] = \int_{\cM} d^4x\,\sqrt{-g} \left[ R + \Lambda + \frac{1}{2}g^{\mu\nu} (\del_{\mu}\varphi)(\del_{\nu}\varphi) + \frac{1}{2} m^2\varphi^2 + \Lambda\varphi \right].
$$
Note that the Minkowski metric $\eta_{\mu\nu}$ is not a solution of this theory, due to the presence of the cosmological constant. Nevertheless, the action is an off-shell object, and we can evaluate it on any background, not just on the solutions of the theory. In particular, putting $g_{\mu\nu} = \eta_{\mu\nu}$ gives:
$$
S[\eta,\varphi] = \int_{\cM} d^4x \left[ \Lambda + \frac{1}{2}\eta^{\mu\nu} (\del_{\mu}\varphi)(\del_{\nu}\varphi) + \frac{1}{2} m^2\varphi^2 + \Lambda\varphi \right].
$$
The standalone cosmological constant in the first term can be ignored, since that term does not depend on matter fields and is therefore just an additive constant in the otherwise matter-only action. The remainder of the action is almost exactly the same as the scalar field action in special relativity --- only the presence of the last term $\Lambda\varphi$ makes a difference. As a consequence of this, we conclude that the equivalence principle is violated in the original theory by the presence of this term.

Now we apply the same test to the $\itGamma_{\lc}$ actions in (\ref{RazniIzboriZaGamaNulaZaNoviVerteks}). Consider the configuration of the background fields which corresponds to nonzero matter fields $\phi$ in flat Minkowski spacetime $M_4$ described by $j$ and $\vec{n}$. Taking $\cM = M_4$, one can construct its triangulation $T(M_4)$, and introduce edge-lengths $L_{\epsilon}$ as in the usual classical Regge gravity. From those one can compute the values of the background fields $j(L)$ and $\vec{n}(L)$ which correspond to flat Minkowski spacetime. Now we evaluate $\itGamma_{\lc}$ on this background,
$$
\itGamma_{\lc}(\phi) \equiv \itGamma_{\lc} (j(L),\phi) = \gamma \sum_v \lc_v S_v(j(L)) + \sum_v \lc_v S_v^{\rm matter}(j(L),\phi) ,
$$
where we consider edge-lengths $L$ to be fixed, so that the effective action is a functional of $\phi$ only. We are allowed to evaluate this action on the Minkowski background because the action is {\em off-shell}, and the flat background geometry does not need to satisfy the equations of motion for this action. The evaluation procedure corresponds to the approximation where the gravitational degrees of freedom are ``frozen-out'', and the backreaction of matter on geometry is neglected, as is the case in special relativity. Consequently, the first sum on the right-hand side is constant and can be dropped from the effective action, which then reads
$$
\itGamma_{\lc}(\phi) = \sum_v \lc_v S_v^{\rm matter}(j(L),\phi) .
$$
Now, using the equivalence principle, we can identify this effective action to the action for the matter fields in flat Minkowski spacetime, $S_{\rm Minkowski}(\phi)$, so we conclude that
$$
S_{\rm Minkowski}(\phi) = \sum_v \lc_v S_v^{\rm matter}(j(L),\phi).
$$
However, if we perform the same procedure to the effective action $\itGamma_+$ in (\ref{RazniIzboriZaGamaNulaZaNoviVerteks}), we obtain a different expression for the Minkowski action,
$$
S_{\rm Minkowski}(\phi) = S_M(j(L),\phi) \equiv \sum_v S_v^{\rm matter}(j(L),\phi).
$$
Since the two expressions for $S_{\rm Minkowski}(\phi)$ are not equal, one of the effective actions, $\itGamma_+$ or $\itGamma_{\lc}$, gives the wrong flat spacetime action for the matter fields, and thus violates the equivalence principle.

Regarding the above procedure, it is important to stress that we are allowed to perform it despite the fact that the flat Minkowski spacetime might not be a solution of the equations of motion for the action $\itGamma_{\lc}$. Namely, as has been explained in Appendix \ref{AppendixC} and stressed throughout the paper, all actions in (\ref{RazniIzboriZaGamaNulaZaNoviVerteks}) have been calculated {\em off-shell}, and we may evaluate them on an arbitrary configuration of background fields (as long as the background corresponds to a Regge-like geometry).

On the other hand, it should be remarked that, since the action is evaluated on an off-shell background in the gravitational sector, the resulting effective action for matter fields is derived under the assumption of the nonzero source term in the gravitational sector. In that light, it is not entirely clear whether the equivalence principle is supposed to hold in such a situation, and our conclusions regarding the importance of its violation may be questioned. This is a conceptual issue which requires further investigation.

Next, note that the matter coupling of the form (\ref{EPRLFKverteksAmplitudaSaMaterijom}) does not have the above problem with the equivalence principle. Namely, the matter sector of all effective actions in (\ref{RazniIzboriZaGamaNula}) is the same --- so when evaluated on Minkowski spacetime, all effective actions give the same classical action for the matter fields. Therefore, the matter coupling of the form (\ref{EPRLFKverteksAmplitudaSaMaterijom}) is compatible with the equivalence principle for all effective actions in the theory, while the matter coupling of the form (\ref{NovoKuplovanjeMaterije}) necessarily violates the equivalence principle for all but one effective action in the theory. This is the main reason why we have preferred (\ref{EPRLFKverteksAmplitudaSaMaterijom}) over (\ref{NovoKuplovanjeMaterije}) in this paper.

Thus, on the upside, the coupling of matter to gravity of the form (\ref{NovoKuplovanjeMaterije}) does indeed resolve the problem of antigravity. On the downside, it still fails to resolve the full problem of multiple classical limits, since the ``mixed'' effective actions $\itGamma_{\lc}$ are still present in the theory. Moreover, the equivalence principle is violated for those ``mixed'' effective actions. In order to resolve this problem, one needs to somehow eliminate $\itGamma_{\lc}$ from the set of solutions (\ref{RazniIzboriZaGamaNulaZaNoviVerteks}), and the matter coupling of the form (\ref{NovoKuplovanjeMaterije}) does not help one do that. Finally, another downside is that the vertex amplitude $A_v$ with asymptotics (\ref{NovoKuplovanjeMaterije}) is not easy to construct, if possible at all.

\section{\label{SectionConclusions}Conclusions}

In this paper we were investigating the classical limit of the EPRL/FK spinfoam model of quantum gravity, with the coupling of matter fields. In particular, we have shown that the presence of two conjugate exponents in the large-spin asymptotics of the EPRL/FK vertex amplitude (the so called {\em cosine problem}) gives rise to two classical limits of the theory --- one describing classical theory of gravity, and the other describing the same theory but with an opposite coupling constant, i.e. antigravity. The presence of antigravity can be established only once the matter fields are coupled to the spinfoam model, since the attractive or repulsive nature of the gravitational interaction depends on the relative sign of the gravitational and matter sectors in the effective action. Antigravity obviously represents a problem for the original formulation of the theory, since it follows that the unique classical limit of the theory does not exist.

After the introduction given in section \ref{SectionIntroduction}, in section \ref{SectionEPRLFKmodel} we gave a short introduction to the EPRL/FK spin foam model, discussed the results for the asymptotics of the vertex amplitude, and generalized the model to include matter fields. In section \ref{SectionEffectiveAction} we have introduced the background field method for evaluating the effective action in quantum field theory. After discussing the possibilities of having multiple classical limits in the same theory, we have adapted the method to the spin foam setting, in order to evaluate the effective action for the EPRL/FK model with matter in the classical limit. The computation was performed in section \ref{SectionClassicalLimit}, with several conclusions. First, it was found that the model has multiple classical limits, depending on what kind of geometry was described by the choice of the configuration of the background fields. One of these limits indeed corresponds to classical general relativity. However, since matter fields are present in the model, we have also discovered that there is not one, but rather {\em a whole class} of classical limits, for the same choice of the background fields. In addition to ordinary gravity, one of these classical limits corresponds to antigravity, while other choices ``interpolate'' between the gravitational and antigravitational limits. This represents a problem for the theory, which was discussed in  section \ref{SectAntigravity}. After providing the proper interpretation of various classical limits, we have established that the reason for the appearance of these extraneous limits is the presence of two exponents in the asymptotics of the vertex amplitude.

Three strategies which attempt to resolve the problem of these extraneous classical limits have been discussed. Arguably the most obvious one is to redefine the vertex amplitude so that it does not have the second exponent in the large spin asymptotics. This can certainly be done and several examples were given. However, there is no obvious unique way of fixing the new definition of the vertex amplitude. This in turn leads to having a whole class of theories, all of which have the correct classical limit, but different quantum behavior.

The second strategy deals with the attempt to use the boundary wavefunction formalism in order to suppress the extraneous classical limits. However, this approach is of limited success at best. Namely, there are various conceptual issues about the meaning of putting boundary conditions on the effective action, and there are technical issues about the boundary wavefunction satisfying the Hamiltonian constraint. But most importantly, there is no clean way to include the boundary wavefunction in the set of physical solutions of the theory, while keeping the conjugate boundary wavefunction outside of this set. The simple analysis shows that the theory which implements this requirement also violates unitarity, since it requires that the Hamiltonian of the theory must not be self-adjoint. This is unappealing from the point of view of quantum mechanics.

The third strategy deals with a more intricate way one could couple matter fields to the EPRL/FK spinfoam model, in order to avoid the relative sign flip between gravity and matter in the conjugate exponent, and thus the appearance of antigravity. While this could in principle be made to work, not all extraneous classical limits can be eliminated from the theory, which leads not only to multiple classical limits for gravity, but for matter fields as well. This in turn indicates the presence of multiple classical limits for matter fields in flat Minkowski spacetime, in contradiction with both the equivalence principle and the experimental fact that we can observe only one of these limits. Such a bad situation is a clear consequence of the assumed complicated coupling of matter to gravity. In general, these strategies of intricate coupling of matter to gravity are sensitive to the validity of the equivalence principle, and special care must be taken not to violate it.

As a final remark, one can discuss which of these three strategies would be best to pursue forward, or maybe formulate yet another way to deal with the cosine problem. While all options are open, we believe that the first strategy is the most conservative choice, as it preserves the known physics --- it does not violate either unitarity nor the equivalence principle. The nonuniqueness problem is of course a severe one, but nonetheless also implicit in other approaches as well.

\begin{acknowledgments}
The author would like to thank Aleksandar Mikovi\'c for helpful discussion and suggestions, , and to the anonymous Referee for remarks that have improved the quality of the presentation. This work has been supported by the FCT grant SFRH/BPD/46376/2008 and the FCT project PEst-OE/MAT/UI0208/2011.
\end{acknowledgments}

\appendix

\section{\label{AppendixA}Notation and conventions}

When discussing a $2$-complex, we denote its vertices, edges and faces as $v$, $e$ and $f$, respectively. We use the $\in$ symbol to denote adjacency. For example, the sum
$$
\sum_{f\in v} F(v,f)
$$
denotes the sum over all faces that are connected to a given vertex. Similarly,
$$
\sum_{v\in f} F(v,f)
$$
denotes the sum over all vertices that are connected to a given face. Note that if $v\in f$, then $f\in v$ as well.

\bigskip

The ``big-O'' and ``small-o'' symbols are defined in the usual way. One says that $f(x)=O(g(x))$ iff
$$
\limsup_{x\to\infty} \frac{\left| f(x) \right|}{\left| g(x) \right|} < \infty.
$$
Also, one says that $f(x) = o(g(x))$ iff
$$
\lim_{x\to\infty} \frac{\left| f(x) \right|}{\left| g(x) \right|} =0.
$$
The usage of these symbols is convenient since it simplifies notation, and allows one to talk about the scaling of various quantities in the large-spin limit $j\to\infty$ without explicitly introducing a parameter to be taken to infinity.

\section{\label{AppendixB}Solving the effective action equation}

In order to demonstrate the background field method for calculating the effective action outlined in section \ref{SectionEffectiveAction}, we shall perform the explicit procedure of solving equation (\ref{DefinicijaEffDejstvaUtp}) up to the first two orders. For simplicity, we shall discuss the mechanical system with one degree of freedom, rather than field theory. Therefore, the action $S[\phi]$ will become an ordinary real-valued function $S(x)$ over $x\in\realni$, while the path integral will become an ordinary integral over the set of real numbers.

Equation (\ref{DefinicijaEffDejstvaUtp}) can then be rewritten as an ordinary integrodifferential equation
\begin{equation} \label{AppAPocetnaJednacina}
e^{i\itGamma(x)} = \int_{\realni} dy \; e^{ i S(x+y) - i\itGamma'(x)y },
\end{equation}
where the prime denotes the derivative with respect to $x$. We are looking for an asymptotic solution of this equation in the limit $x\to\infty$, assuming that  $S(x) = O(x)$. First we write the effective action in the form
$$
\itGamma(x) = \itGamma_0(x) + \itGamma_1(x) + o(\itGamma_1),
$$
where
\begin{equation} \label{AppAOrdersOfGamma}
\itGamma_0 = O(x), \qquad \itGamma_1 = o(\itGamma_0) = o(x).
\end{equation}
Substituting this into (\ref{AppAPocetnaJednacina}) and expanding $S(x+y)$ into power series around the point $x$, we obtain
\begin{equation} \label{AppUsputnaJnaZaEffDejstvo}
e^{i\itGamma_0 + i\itGamma_1} = e^{iS} \int_{\realni} dy \; e^{ i (S' - \itGamma'_0 - \itGamma'_1) y + \frac{i}{2}S'' y^2 + \frac{i}{6}S''' y^3 + \dots}.
\end{equation}
Note the order of various terms:
$$
\itGamma_0 = O(x), \qquad S = O(x), \qquad \itGamma_1 = o(x),
$$
$$
\itGamma'_0 = O(1), \qquad S' = O(1), \qquad \itGamma'_1 = o(1),
$$
$$
S'' = O(x^{-1}), \qquad S''' = O(x^{-2}), \qquad \dots
$$
The cubic and higher terms in the exponent can be neglected, since corresponding coefficients scale as $O(x^{-2})$ and lower. We can then rewrite the exponent on the right-hand side as
$$
 i (S' - \itGamma'_0)y - i \itGamma'_1 y + \frac{i}{2}S'' y^2,
$$
where the three terms are of orders $O(1)$, $o(1)$ and $O(x^{-1})$, respectively. Note that if the $O(1)$ term $S'-\itGamma'_0$ does not vanish, the integrand is dominated by an oscillatory term, and consequently the integral is exponentially suppressed. The integral will be nonzero only if we set $S' = \itGamma'_0$, which gives
\begin{equation} \label{AppRnjeZaGamaNula}
\itGamma_0(x) = S(x) + const,
\end{equation}
where the arbitrary constant can be neglected because it does not contribute to the effective equations of motion. The equation (\ref{AppUsputnaJnaZaEffDejstvo}) then reduces to
$$
e^{i\itGamma_1} = \int_{\realni} dy \; e^{ \frac{i}{2}S'' y^2 - i \itGamma'_1 y},
$$
where the exponent can now be represented in the following form by completing the square:
$$
\frac{1}{2}S'' y^2 - \itGamma'_1 y  = \frac{S''}{2} \left( y - \frac{\itGamma'_1}{S''} \right)^2 - \frac{(\itGamma'_1)^2}{2S''}.
$$
At this point the integral can be evaluated, and we obtain
$$
e^{i\itGamma_1} =  e^{-i\frac{(\itGamma'_1)^2}{2S''}} \left[ \sqrt{\frac{2\pi}{|S''|}} \; e^{i\frac{\pi}{4}\sgn{(S'')}} \right],
$$
which can be rearranged in the form
\begin{equation} \label{AppDifJnaZaGamaJedan}
\itGamma_1 = \frac{i}{2} \ln |S''| + \frac{\pi}{4}\sgn(S'') - \frac{i}{2}\ln(2\pi) -\frac{(\itGamma'_1)^2}{2S''}.
\end{equation}
This is a first-order differential equation for $\itGamma_1(x)$. However, note that the first term on the right is leading as $O(\log x)$, while the remaining three terms are subleading and can therefore be neglected, in particular the term containing $\itGamma'_1$. Consequently we have
\begin{equation} \label{AppRnjeZaGamaJedan}
\itGamma_1 = \frac{i}{2} \ln |S''| = O(\log x) = o(x).
\end{equation}
As a verification of the procedure, note that
$$
\itGamma'_1 = \frac{i}{2} \frac{S'''}{S''} = O(x^{-1}) = o(1),
$$
as we have assumed from the beginning. In particular, the last term in (\ref{AppDifJnaZaGamaJedan}) is of order $O(x^{-1}) = o(\log x)$, as expected.

Putting together the obtained results (\ref{AppRnjeZaGamaNula}) and (\ref{AppRnjeZaGamaJedan}) for $\itGamma_0$ and $\itGamma_1$ we obtain the complex effective action
$$
\itGamma(x) = S(x) + \frac{i}{2} \ln |S''(x)|,
$$
which can be Wick-rotated according to prescription (\ref{WickRotacija}) to obtain the final familiar form
$$
\itGamma(x) = S(x) + \frac{1}{2} \ln |S''(x)|.
$$

As a final note, it is straightforward to generalize this result for an arbitrary number of degrees of freedom, giving the well-known formula for the semiclassical effective action in field theory,
$$
\itGamma[\phi] = S[\phi] + \frac{1}{2} \tr \log S''[\phi].
$$

\section{\label{AppendixC}Derivation of the effective action equation}

In quantum field theory, one introduces the concept of the {\em effective action} in the following way. Starting from the path integral state-sum for the classical action $S[\varphi]$,
\begin{equation} \label{AppCpathInt}
Z[J] = \int \cD \varphi e^{iS[\varphi] + i\int \varphi J},
\end{equation}
one first defines the energy functional $W[J]$ as the logarithm of the state-sum,
\begin{equation} \label{AppCdefEnFunc}
W[J] = -i \log Z[J],
\end{equation}
which should not be confused with the notation for the EPRL/FK vertex amplitude, despite the letter $W$ used to denote it.

Then one defines the {\em effective action} $\itGamma[\phi]$, a functional of an {\em arbitrarily chosen} field $\phi$ (called the {\em background field}), as a Legendre transformation of $W[J]$,
\begin{equation} \label{AppClezandr}
\itGamma[\phi] = W[J[\phi]] - \int \phi J[\phi].
\end{equation}
Here the source $J$ is written as a functional $J[\phi]$ of the chosen background field $\phi$, since for every choice of $\phi$ we should have a corresponding choice of $J$. This choice can be determined by taking the derivative of (\ref{AppClezandr}),
\begin{equation} \label{AppClezandrDiferenc}
\frac{\delta \itGamma[\phi]}{\delta\phi} = \int \frac{\delta W[J]}{\delta J}\Big|_{J[\phi]} \frac{\delta J[\phi]}{\delta \phi} - J[\phi] - \int \phi \frac{\delta J[\phi]}{\delta\phi}.
\end{equation}
and requiring that the first term on the right-hand side cancels the third term. This will happen if
\begin{equation} \label{AppCjnaZaJ}
\frac{\delta W[J]}{\delta J}\Big|_{J[\phi]} = \phi,
\end{equation}
which gives us an implicit definition of the functional $J[\phi]$ for any given background $\phi$. It is important to note that this equation determines $J[\phi]$, rather than $\phi$ itself, which is completely arbitrary. After the cancellation (\ref{AppClezandrDiferenc}) reduces to
\begin{equation} \label{AppCrnjeZaJ}
\frac{\delta \itGamma[\phi]}{\delta\phi} = - J[\phi].
\end{equation}

The effective action in the form (\ref{AppClezandr}) is not very useful in practice, since in order to evaluate it one needs to calculate $W[J]$ from $Z[J]$, then invert the equation (\ref{AppCjnaZaJ}) in order to solve it for $J[\phi]$, and finally put all that into (\ref{AppClezandr}) and evaluate $\itGamma[\phi]$. However, there is a more clever method, which expresses $\itGamma[\phi]$ directly in terms of the classical action $S[\varphi]$. That is the equation (\ref{DefinicijaEffDejstvaUtp}) used in the main text. In order to derive it, we begin by taking the exponent of (\ref{AppClezandr}),
$$
\exp \left(i\itGamma[\phi] \right) = \exp \left( iW[J[\phi]] \right) \exp \left( -i\int \phi J[\phi] \right),
$$
and we use the definition (\ref{AppCdefEnFunc}) to rewrite this in terms of the state sum,
\begin{equation} \label{AppCuspJna}
\exp \left(i\itGamma[\phi] \right) = Z[J[\phi]] \exp \left( -i\int \phi J[\phi] \right).
\end{equation}
Next we want to get rid of the other exponent on the right-hand side. To this end, we shift the integration variable $\varphi$ in the path integral (\ref{AppCpathInt}) by $\phi$,
$$
\varphi = \phi + \tilde{\phi},
$$
where $\tilde{\phi}$ is the new integration variable and $\phi$ is the fixed background. Given that the background $\phi$ is fixed, we have $\cD\varphi = \cD \tilde{\phi}$. The state sum is then rewritten as
$$
Z[J] = \exp\left( i\int \phi J \right) \int \cD \tilde{\phi} e^{iS[\phi+\tilde{\phi}] + i\int \tilde{\phi} J}.
$$
Evaluating the state sum on $J[\phi]$ and substituting into (\ref{AppCuspJna}), the exponent in front of the path integral cancels the exponent in (\ref{AppCuspJna}), and we obtain
$$
\exp \left(i\itGamma[\phi] \right) = \int \cD \tilde{\phi} e^{iS[\phi+\tilde{\phi}] + i\int \tilde{\phi} J[\phi]} .
$$
As a final step, we eliminate the remaining term $J[\phi]$ in the path integral by using (\ref{AppCrnjeZaJ}) to obtain
\begin{equation} \label{AppCrezultat}
e^{i\itGamma[\phi]} = \int \cD \tilde{\phi} e^{iS[\phi+\tilde{\phi}] - i\int \frac{\delta \itGamma[\phi]}{\delta\phi} \tilde{\phi} } .
\end{equation}
This is the desired equation (\ref{DefinicijaEffDejstvaUtp}) used in the main text. It is a functional integrodifferential equation for the effective action $\itGamma$ evaluated on an arbitrary background field $\phi$, expressed in terms of the path integral and the classical action $S$.

An important remark is in order. Namely, the background field $\phi$ is kept {\em completely arbitrary} throughout the derivation process. In particular, it is not assumed that it satisfies any kind of equations of motion. In other words, the above derivation of (\ref{AppCrezultat}) is a fully off-shell and nonperturbative calculation, and the resulting equation (\ref{AppCrezultat}) itself is valid in general.

In contrast to this general formula, note that the effective action is commonly introduced on-shell --- with an assumption that the background field $\phi$ satisfies the effective equations of motion (compare with (\ref{AppCrnjeZaJ})),
$$
\frac{\delta \itGamma[\phi]}{\delta\phi} = 0.
$$
If this assumption is satisfied, the second term in the exponent on the right-hand side of (\ref{AppCrezultat}) vanishes, and our functional integrodifferential equation then drastically simplifies to the form
\begin{equation} \label{AppCrezultatOnShell}
e^{i\itGamma[\phi]} = \int \cD \tilde{\phi} e^{iS[\phi+\tilde{\phi}] } .
\end{equation}
In contrast to (\ref{AppCrezultat}), equation (\ref{AppCrezultatOnShell}) is already solved for the effective action $\itGamma$, and one only needs to evaluate the remaining path integral in order to find an explicit expression for $\itGamma[\phi]$. Given that this is a much simpler from the calculational point of view, it is often a common practice to discuss only equation (\ref{AppCrezultatOnShell}). However, the price for this apparent simplicity is the requirement that the background $\phi$ then must be a solution of the effective equations of motion, rather than completely arbitrary. This can be very problematic if the form of the effective action which determines these equations of motion is not already known. Namely, in order to determine the effective action using (\ref{AppCrezultatOnShell}), one needs to know in advance the solution of its equations of motion, which is in most cases a circular problem. Therefore equation (\ref{AppCrezultatOnShell}) can be of limited applicability. In contrast, the general equation (\ref{AppCrezultat}) is valid for an arbitrary background $\phi$, and can be solved systematically by an iterative procedure, as described in section \ref{SectionEffectiveAction} and Appendix \ref{AppendixB}.

\end{document}